\documentstyle[11pt,amssymb,epsf]{article}

\def\ben{\begin{equation}}
\def\een{\end{equation}}

\let\a=\alpha    
    
  \let\n=\nu

\let\C=\Chi

\def\nn{\nonumber} \def\bd{\begin{document}} \def\ed{\end{document}}
\def\ds{\documentstyle} \let\fr=\frac \let\bl=\bigl \let\br=\bigr
\let\Br=\Bigr \let\Bl=\Bigl
\let\bm=\bibitem
\let\na=\nabla
\let\pa=\partial \let\ov=\overline
\newcommand{\be}{\begin{equation}}
\newcommand{\ee}{\end{equation}}
\def\ba{\begin{array}}
\def\ea{\end{array}}
\def\ft#1#2{{\textstyle{{\scriptstyle #1}\over {\scriptstyle #2}}}}
\def\fft#1#2{{#1 \over #2}}
\def\del{\partial}
\def\vp{\varphi}
\def\sst#1{{\scriptscriptstyle #1}}
\def\oneone{\rlap 1\mkern4mu{\rm l}}
\def\td{\tilde}
\def\wtd{\widetilde}
\def\ie{\rm i.e.\ }
\def\dalemb#1#2{{\vbox{\hrule height .#2pt
        \hbox{\vrule width.#2pt height#1pt \kern#1pt
                \vrule width.#2pt}
        \hrule height.#2pt}}}
\def\square{\mathord{\dalemb{6.8}{7}\hbox{\hskip1pt}}}
\newcommand{\ho}[1]{$\, ^{#1}$}
\newcommand{\hoch}[1]{$\, ^{#1}$}
\newcommand{\bea}{\begin{eqnarray}}
\newcommand{\eea}{\end{eqnarray}}
\newcommand{\ra}{\rightarrow}
\newcommand{\lra}{\longrightarrow}
\newcommand{\Lra}{\Leftrightarrow}
\newcommand{\ap}{\alpha^\prime}
\newcommand{\bp}{\tilde \beta^\prime}
\newcommand{\tr}{{\rm tr} }
\newcommand{\Tr}{{\rm Tr} }
\def\0{{\sst{(0)}}}
\def\1{{\sst{(1)}}}
\def\2{{\sst{(2)}}}
\def\3{{\sst{(3)}}}
\def\4{{\sst{(4)}}}
\def\5{{\sst{(5)}}}
\def\6{{\sst{(6)}}}
\def\7{{\sst{(7)}}}
\def\8{{\sst{(8)}}}
\def\n{{\sst{(n)}}}
\def\cA{{{\cal A}}}
\def\cF{{{\cal F}}}
\def\tV{\widetilde V}
\def\tW{\widetilde W}
\def\tH{\widetilde H}
\def\tE{\widetilde E}
\def\tF{\widetilde F}
\def\tA{\widetilde A}
\def\im{{{\rm i}}}
\def\tY{{{\wtd Y}}}
\def\ep{{\epsilon}}
\def\vep{{\varepsilon}}
\def\R{\rlap{\rm I}\mkern3mu{\rm R}}
\def\bD{{{\bar D}}}

\def\R{\rlap{\rm I}\mkern3mu{\rm R}}
\def\bD{{{\bar D}}}
\def\R{{{\Bbb R}}}
\def\C{{{\Bbb C}}}
\def\H{{{\Bbb H}}}
\def\CP{{{\Bbb C}{\Bbb P}}}
\def\RP{{{\Bbb R}{\Bbb P}}}
\def\Z{{{\Bbb Z}}}
\def\bA{{{\Bbb A}}}
\def\bB{{{\Bbb B}}}
\def\bC{{{\Bbb C}}}
\def\bZ{{{\Bbb Z}}}

\def\cosec{{\,\hbox{cosec}\,}}

\newcommand{\tamphys}{\it Center for Theoretical Physics,
Texas A\&M University, College Station, TX 77843, USA}
\newcommand{\umich}{\it Michigan Center for Theoretical Physics,
University of Michigan\\ Ann Arbor, MI 48109, USA}
\newcommand{\upenn}{\it Department of Physics and Astronomy,
University of Pennsylvania\\ Philadelphia,  PA 19104, USA}
\newcommand{\SISSA}{\it  SISSA-ISAS and INFN, Sezione di Trieste\\
Via Beirut 2-4, I-34013, Trieste, Italy}

\newcommand{\ihp}{\it Institut Henri Poincar\'e\\
  11 rue Pierre et Marie Curie, F 75231 Paris Cedex 05}

\newcommand{\damtp}{\it DAMTP, Centre for Mathematical Sciences,
 Cambridge University\\ Wilberforce Road, Cambridge CB3 OWA, UK}
\newcommand{\itp}{\it Institute for Theoretical Physics, University of
California\\ Santa Barbara, CA 93106, USA}

\newcommand{\auth}{M. Cveti\v{c}\hoch{\dagger}, G.W. Gibbons\hoch{\sharp},
H. L\"u\hoch{\star} and C.N. Pope\hoch{\ddagger}}

\begin{document}
\begin{flushright}
\hfill{DAMTP-2001-101}\ \ \ {CTP TAMU-31/01}\ \ \ {UPR-966-T}\ \ \
{MCTP-01-55}\\
{November 2001}\ \ \
{hep-th/0111096}
\end{flushright}


\begin{center}
{ \large {\bf Orientifolds and Slumps in $G_2$ and Spin(7) Metrics}}

\vspace{5pt}
\auth

\vspace{3pt}
{\hoch{\dagger}\upenn}

\vspace{3pt}


\vspace{3pt}
{\hoch{\sharp}\damtp}

\vspace{3pt}
{\hoch{\star}\umich}

\vspace{3pt}
{\hoch{\ddagger}\tamphys}

\vspace{3pt}

\underline{ABSTRACT}
\end{center}

   We discuss some new metrics of special holonomy, and their roles in
string theory and M-theory.  First we consider Spin(7) metrics denoted
by $\bC_8$, which are complete on a complex line bundle over $\CP^3$.
The principal orbits are $S^7$, described as a triaxially squashed
$S^3$ bundle over $S^4$.  The behaviour in the $S^3$ directions is
similar to that in the Atiyah-Hitchin metric, and we show how this
leads to an M-theory interpretation with orientifold D6-branes wrapped
over $S^4$.  We then consider new $G_2$ metrics which we denote by
$\bC_7$, which are complete on an $\R^2$ bundle over $T^{1,1}$, with
principal orbits that are $S^3\times S^3$.  We study the $\bC_7$
metrics using numerical methods, and we find that they have the
remarkable property of admitting a $U(1)$ Killing vector whose length
is nowhere zero or infinite.  This allows one to make an everywhere
non-singular reduction of an M-theory solution to give a solution of
the type IIA theory.  The solution has two non-trivial $S^2$ cycles,
and both carry magnetic charge with respect to the R-R vector field.
We also discuss some four-dimensional hyper-K\"ahler metrics described
recently by Cherkis and Kapustin, following earlier work by
Kronheimer.  We show that in certain cases these metrics, whose
explicit form is known only asymptotically, can be related to metrics
characterised by solutions of the $su(\infty)$ Toda equation, which
can provide a way of studying their interior structure.

\pagebreak
\setcounter{page}{1}

\tableofcontents
\addtocontents{toc}{\protect\setcounter{tocdepth}{3}}
\vfill\eject

\section{Introduction}

   Metrics of special holonomy have played an important role in many
areas of string theory and M-theory.  Compact metrics have been used
extensively for Kaluza-Klein dimensional reduction to four, and other
dimensions.  Non-compact metrics of special holonomy have been used as
generalisations of the usual flat metric on the space transverse to a
$p$-brane, providing configurations with lesser supersymmetry, and by
turning on additional fluxes, configurations that break other
symmetries, such as the conformal symmetry in the usual AdS/CFT
correspondence.  In this paper, we shall investigate several examples
involving new metrics with special holonomy.

   Our first example is a family of eight-dimensional metrics of
cohomogeneity one and Spin(7) holonomy, defined on a certain complex
line bundle over $\CP^3$.  The principal orbits in these metrics are
$S^7$, viewed as a triaxially-squashed $S^3$ bundle over $S^4$.  The
system of four first-order equations following from requiring Spin(7)
holonomy were obtained recently in \cite{hitch,g2spin7}, and a
numerical analysis in \cite{g2spin7} indicated that complete
asymptotically local conical (ALC) metrics, denoted by $\bC_8$ there,
arise as solutions of these equations.  There is a non-trivial
parameter, characterising the degree of squashing of the $\CP^3$ bolt.
These solutions were found to have a behaviour similar to that seen in
the four-dimensional Atiyah-Hitchin metric \cite{atihit,gibman}, with
one direction on the $S^3$ fibres collapsing on the singular orbit at
short distance, whilst a different direction in the $S^3$ stabilises
to a constant radius at large distance \cite{g2spin7}.

    If one takes the product of the Atiyah-Hitchin metric and 
seven-dimensional Minkowski spacetime, one obtains a solution of
$D=11$ supergravity which, after reduction on the circle of
asymptotically-stabilised radius, admits an interpretation as a
D6-brane orientifold plane \cite{Sen}.  We show in this paper that an
analogous interpretation can be given to an M-theory solution of the
product of a $\bC_8$ metric with 3-dimensional Minkowski spacetime,
which from the type IIA viewpoint becomes D6-brane orientifold plane
wrapped on $S^4$.  We do this by making a detailed analysis of the
coordinate identifications required by the regularity of the $\bC_8$
metrics, and by making a perturbative analysis of the large-distance
asymptotics, to establish the mass of the solutions.

    Our second example is concerned with seven-dimensional metrics of
cohomogeneity one and $G_2$ holonomy.  A wide class of metrics with
$S^3\times S^3$ principal orbits was studied in
\cite{g2spin7,brgogugu}, with six metric functions characterising
homogeneous deformations of the $S^3\times S^3$, and the first-order
equations implying $G_2$ holonomy were obtained.  In this paper, we
study a class of solutions of these equations in which the principal
$S^3\times S^3$ orbits degenerate to a bolt with the topology of the
5-dimensional space $T^{1,1}$, as a consequence of the collapse of a
circle at short distance.  We obtain short-distance Taylor expansions
for such solutions, and use them in order to integrate numerically to
search for regular metrics.  We find indications that there exist
non-singular ALC metrics with a non-trivial parameter that
characterises the degree of squashing of the $T^{1,1}$ bolt.  At the
upper limit of the range of this parameter, the metric becomes
asymptotically conical (AC).  The ALC metrics, which we shall denote
by $\bC_7$, are asymptotic to the product of a circle and an AC
six-metric on the cone over $T^{1,1}/Z_2$.  The $Z_2$ identification
is a consequence of the requirement of regularity at the degenerate
orbit at short distance.  By contrast, the $T^{1,1}$ of the bolt at
short distance can be modified globally, if desired, to $T^{1,1}/Z_N$
for any integer $N$.  (This does not affect the $T^{1,1}/Z_2$
structure at infinity.)

    An intriguing feature of these new metrics, which are defined on
an $\R^2$ bundle over $T^{1,1}$, is that they admit a circle action
whose length is everywhere finite and non-zero.  This means that one
can perform a completely non-singular Kaluza-Klein reduction on this
circle, from a starting point of an M-theory solution comprising the
$G_2$ metric times four-dimensional Minkowski spacetime.  In ten
dimensions, this gives a solution of the type IIA theory.  
The R-R 2-form charge of the solution is equal to
the integer $N$ characterising the $T^{1,1}/Z_N$ geometry of the bolt.
Because the dilaton stablises everywhere, the large $N$
limit of the solution is a valid approximation at the level of string theory,
both at large and small distances.

   In the remainder of the paper, we study some new four-dimensional
hyper-K\"ahler metrics, which have recently been discussed in
\cite{Cherkis}.\footnote{These metrics were also discussed a while ago
by Kronheimer, but they did not appear explicitly in print.}  Of
course these are not of cohomogeneity one (all such metrics were well
studied and fully classified long ago), and indeed in general the
complete metrics discussed in \cite{Cherkis} would have no continuous
symmetries at all.  The asymptotic forms of the complete metrics are
obtained explicitly in \cite{Cherkis}; these themselves would, if
extended to short distance, lead to singularities, but it is argued
that there exist smooth ``resolutions'' of these singularities.  All
the asymptotic metrics in \cite{Cherkis} admit a $U(1)\times U(1)$
isometry that acts tri-holomorphically, and which would be lost in the
core of the general complete solution.

   In special cases the resolved metrics discussed in \cite{Cherkis}
would still admit an exact $U(1)$ isometry.  This circle action would 
not act tri-holomorphically within the bulk of the solution.  When
such a $U(1)$ isometry occurs, the hyper-K\"ahler metric must
necessarily be describable within a class of metrics introduced in
\cite{boyfin,dasgeg},  in which there is a function of three variables that
satisfies an $su(\infty)$ Toda equation.  In principle, therefore, one
can probe the properties as one moves in from the asymptotic region in
this subclass of the complete metrics described by Cherkis and Kapustin, by 
perturbing around the asymptotic form after re-expressing it in the
Toda-metric variables.   We study this for certain special cases of
the metrics in \cite{Cherkis}.

\section{$\bC_8$ metrics of Spin(7) holonomy, and orientifold planes}

\subsection{The construction of the $\bC_8$ metrics}\label{c8con}

    In \cite{g2spin7}, a numerical analysis was used to demonstrate
the existence of new cohomogeneity one metrics of Spin(7) holonomy
on the chiral spin bundle of $S^4$, which were denoted by $\bC_8$.
The principal orbits are $S^7$, viewed as an $S^3$ bundle over $S^4$
in which the 3-spheres are triaxially squashed.  This leads to a
metric ansatz involving four functions of the radial coordinate, given
by \cite{g2spin7}
\bea
ds_8^2=dt^2 + a_i^2\, R_i^2 + b^2\, P_a^2\,,\label{spin7met}
\eea
where the 1-forms $R_i$ and $P_a$ live in the coset
$S^7=SO(5)/SO(3)$, and satisfy
\bea
dP_0 &=& (R_1+ L_1)\wedge P_1 + (R_2+L_2)\wedge P_2 + (R_3+L_3)\wedge
      P_3\,,\nn\\
dP_1 &=& -(R_1+ L_1)\wedge P_0 - (R_2-L_2)\wedge P_3 + (R_3-L_3)\wedge
      P_2\,,\nn\\
dP_2 &=& (R_1- L_1)\wedge P_3 - (R_2+L_2)\wedge P_0 - (R_3-L_3)\wedge
      P_1\,,\nn\\
dP_3 &=& -(R_1- L_1)\wedge P_2 + (R_2-L_2)\wedge P_1 - (R_3+L_3)\wedge
      P_0\,,\nn\\
dR_1 &=& -2 R_2\wedge R_3 - \ft12 (P_0\wedge P_1 + P_2\wedge
P_3)\,,\nn\\
dR_2 &=& -2 R_3\wedge R_1 - \ft12 (P_0\wedge P_2 + P_3\wedge P_1)\,,\nn\\
dR_3 &=& -2 R_1\wedge R_2 - \ft12 (P_0\wedge P_3 + P_1\wedge P_2)\,.
\label{so5d}
\eea
(The generators $L_{AB}=-L_{BA}$ of $SO(5)$, with $A=0,1,2,3,4$, satisfy
$dL_{AB} = L_{AC}\wedge L_{CB}$, and are
decomposed here as $P_a=L_{a4}$, $R_i=\ft12 (L_{0i} + \ft12
\ep_{ijk}\, L_{jk})$ and $L_i=\ft12 (L_{0i} - \ft12 \ep_{ijk}\,
L_{jk})$, where $i=1,2,3$.)
The first-order equations implying that (\ref{spin7met}) has
Spin(7) holonomy are given by \cite{hitch,g2spin7}
\bea
\dot a_1 &=& \fft{a_1^2 - (a_2-a_3)^2}{a_2\, a_3} -
\fft{a_1^2}{2b^2}\,,\nn\\
\dot a_2 &=& \fft{a_2^2 - (a_3-a_1)^2}{a_3\, a_1} -
\fft{a_2^2}{2b^2}\,,\nn\\
\dot a_3 &=& \fft{a_3^2 - (a_1-a_2)^2}{a_1\, a_2} -
\fft{a_3^2}{2b^2}\,,\nn\\
\dot b &=& \fft{a_1+a_2+a_3}{4b}\,,\label{spin7fo}
\eea

   It is not clear how to obtain an analytical solution for this
system.  For the restriction $a_1=a_2$, the equations were solved
completely in \cite{cglpnew}, leading to new complete and non-singular
asymptotically locally conical (ALC) metrics $\bB_8$ and $\bB_8^\pm$
on the chiral spin bundle of $S^4$, and a new such metric, denoted by
$\bA_8$, on $\R^8$.  Thus one is now interested in looking for
``triaxial'' solutions such that $a_1\ne a_2\ne a_3\ne a_1$. A
numerical analysis was carried out in \cite{g2spin7}, indicating the
existence of new triaxial solutions, giving rise to complete ALC
metrics on a line bundle over $\CP^3$.

    The procedure used in \cite{g2spin7} for obtaining the new solutions
involved first constructing a short-distance solution, in the form of a
Taylor expansion in the neighbourhood of the bolt, or singular orbit.  At this
stage, one makes a choice about what the topology of the singular orbits will
be.  In the present case, regular possibilities can involve a collapse of 
$S^7$, $S^3$ or $S^1$, giving $\R^8$, an $\R^4$ bundle over $S^4$, or an 
$\R^2$ bundle over $\CP^3$ respectively.  In fact, as shown in \cite{g2spin7}, 
the first case leads only to the $\bA_8$ metric, and the second leads only 
to the $\bB_8$ and $\bB_8^\pm$ metrics, and the original AC Spin(7) metric of 
\cite{brysal,gibpagpop}.  The third possibility, with collapsing $S^1$ orbits, 
leads to the new $\bC_8$ metrics, which we shall discuss more fully here.

   First, therefore, one constructs short-distance Taylor expansions for the 
metric functions, under the assumption that just one of the $a_i$ goes 
to zero on the bolt at $t=0$.  Without loss of generality, one may choose 
this to be $a_1$.  The expansions are then given by \cite{g2spin7}
\bea
a_1 &=& 4 t + \fft{(\lambda^4-40\lambda^2 -48)}{12\lambda^2}\, t^3
+\cdots\,,\nn\\
a_2 &=& \lambda + (1-\ft14\lambda^2)\, t +
\fft{(3\lambda^4 -8\lambda^2 +48)}{32\lambda}\, t^2 + \cdots\,,\nn\\
a_3 &=& -\lambda + (1-\ft14\lambda^2)\, t -
\fft{(3\lambda^4 -8\lambda^2 +48)}{32\lambda}\, t^2 + \cdots\,,\nn\\
b &=& 1 + \ft1{16}(12-\lambda^2)\, t^2 +\cdots\,.\label{onbolt}
\eea
The bolt at $t=0$ has the topology of $\CP^3$, viewed as an $S^2$
bundle over $S^4$.  The constant $\lambda$ characterises the scale of
the $S^2$ fibres relative to the $S^4$ base, and thus parameterises
different homogeneous squashed $\CP^3$ metrics.  In \cite{g2spin7} it
was shown numerically that the solution with these initial data will
be regular everywhere, if $\lambda^2\le 4$, with $\lambda^2=4$
corresponding to the previously-known special case of the complex line
bundle over the ``round'' $\CP^3$ with its Fubini-Study metric (specifically,
the fourth power of the Hopf bundle).
Likewise, the metrics with $\lambda^2<4$ also have the same topology of the
complex line bundle over $\CP^3$.  The metric for $\lambda^2=4$ is AC, and has
$SU(4)$ holonomy, but for $\lambda^2<4$ the metrics are ALC, and have
Spin(7) holonomy.  They locally approach the product of a fixed-radius
circle and the metric of $G_2$ holonomy on the cone over $\CP^3$ at
large distance \cite{g2spin7}.

   Specifically, at large proper distance $t$ the solution approaches
$a_1=a_2\sim t$, $a_3\rightarrow \hbox{const.}$ and $b\sim t$.  The
fact that $a_1$ and $a_2$ become equal at large $t$ implies that the
solutions at large distance can be described as perturbations around
the solutions with $a_1=a_2$ that were obtained previously in
\cite{cglpnew}.  The techniques for doing this are very similar to
ones that can be employed for studying the asymptotic behaviour of the
Atiyah-Hitchin metric.  In that case, it can be viewed as a
perturbation around the Taub-NUT metric.  In order to set the scene
for our analysis of the asymptotic behaviour of the $\C_8$ metrics, it
is therefore useful to carry out such an analysis for the
Atiyah-Hitchin system.  This is done in appendix \ref{ahasymp}.  Of
course the results for this case are already known
\cite{atihit,gibman}, and in that case can be seen directly by
studying the asymptotic properties of the elliptic functions that
appear in the explicit solution.  Our approach, which does not require
knowledge of the explicit form for the solution, is rather simple and
illuminating, and is appropriate for our discussion for the $\bC_8$
metrics, where no explicit solutions are known.

\subsection{Asymptotic behaviour of the $\bC_8$ metrics}

   Here, we study the asymptotic behaviour of the 
$\bC_8$ metrics of Spin(7) holonomy, given by
the ansatz (\ref{spin7met}) with the metric functions $a_i$ and $b$
satisfying (\ref{spin7fo}), which were found numerically in \cite{g2spin7}
and are described in section \ref{c8con}.  

   Proceeding in the same vein as our discussion in appendix
\ref{ahasymp}, we note that the metric functions $a_1$ and $a_2$
become equal asymptotically in the $\bC_8$ solutions, implying that we
can use the already-known exact solutions with $a_1=a_2$ as starting
points for perturbative analyses as large distance.  These exact
solutions, found in \cite{cglpnew} and called $\bA_8$, $\bB_8$ and
$\bB_8^\pm$ there, will play the same role, as the starting point for
a perturbative large-distance analysis, as the Taub-NUT metric did
for the perturbative analysis of the Atiyah-Hitchin metric 
in appendix \ref{ahasymp}.

   We are therefore led to make the following perturbative expansion
at large distance, with
\bea
a_1=2A_0 + A_1\,,\quad a_2=2A_0 + A_2\,,\quad
a_3=2B_0 + A_3\,,\quad b= C_0 +  B\,,\label{orderexp}
\eea
where functions with the subscript 0 denote the unperturbed solutions
in \cite{cglpnew}, while the functions with the subscript 1 denote
linearised perturbations.  (The notation for the upper-case functions
is chosen to fit with the symbols used in \cite{cglpnew}.)
We find that $A_0, B_0$ and $C_0$ satisfy the first-order equations
\be
\dot A_0 = 1-\fft{B_0}{2A_0} - \fft{A_0^2}{C_0^2}\,,\qquad
\dot B_0 =\fft{B_0^2}{2A_0^2} - \fft{B_0^2}{C_0^2}\,,\qquad
\dot C_0 =\fft{A_0}{C_0} + \fft{B_0}{2C_0}\,,\label{zeroeq}
\ee
as in \cite{cglpnew}.

   To solve the equations for the perturbations $A_i$ and $B$, it is
useful to introduce a coordinate gauge function $h$ such that $dt = h
\, dr$, where we also expand $h=h_0 + h_1$. We can substitute
(\ref{orderexp}) into (\ref{spin7fo}), thereby obtaining linearised
equations for $A_i$ and $B$, given by
\bea
A_1' &=& A_1\, \Big(\fft{h_0}{B_0} - \fft{2A_0\,h_0}{C_0^2}\Big) -
             A_2\, \Big(\fft{h_0}{B_0} - \fft{B_0\, h_0}{2A_0^2}\Big)
             -\fft{A_3\, h_0}{2A_0}
             +\fft{4A_0^2\,h_0\, B}{C_0^3} +
             2A_0'\, h_1\,,\nn\\
A_2' &=& -A_1\, \Big(\fft{h_0}{B_0} -\fft{B_0\, h_0}{2A_0^2}\Big) +
             A_2\, \Big(\fft{h_0}{B_0} - \fft{2A_0\,h_0}{C_0^2}\Big) -
             \fft{A_3\, h_0}{2A_0}
            +\fft{4A_0^2\,h_0\, B}{C_0^3} + 2 A_0'\, h_1\,,\nn\\
A_3' &=& -\fft{A_1\,h_0\, B_0^2}{2A_0^3} -
              \fft{A_2\,h_0\, B_0^2}{2A_0^3}
           +A_3\,h_0\, \Big(\fft{B_0}{A_0^2} - \fft{2B_0}{C_0^2}\Big) +
           \fft{4B\,h_0\, B_0^2}{C_0^3} + 2 B_0'\, h_1\,,\nn\\
B' &=& \fft{h_0}{C_0}\, (A_1 + A_2 + A_3) -
       \fft{B\,h_0}{C_0^2}\, (A_0+ B_0) + C_0'\, h_1\,,
\eea
where a prime denotes $d/dr$.

    It is useful to choose a gauge for $h$ such that $A_2=-A_1$.
This requires that
\be
h_1 =\fft{(8A_0\, B- A_3 C_0^3)\, h_0}{2C_0\, (2A_0^3 -2A_0\, C_0^2
+ B_0\, C_0^2)}\,.
\ee
Then, the system admits one simple solution, with
\be
A_3=0\,,\qquad B=0\,,\qquad h_1=0\,,
\ee
and hence $A_1$ satisfies the rather simple equation
\be
2A_1'=  A_1\, h_0\, \Big(\fft{4}{B_0} - \fft{B_0}{A_0^2} -
\fft{4A_0}{C_0^2}\Big)\,.
\ee
Thus we have
\be
A_1=\exp\Big[\ft12\int h_0\, \Big(\fft4{B_0} - \fft{B_0}{A_0^2}
-\ft{4A_0}{C_0^2}\Big)\Big]\,.\label{A1sol}
\ee

   The general solution to the zeroth-order equations (\ref{zeroeq})
was found in \cite{cglpnew}.  This gave rise to an isolated regular
metric, denoted by $\bA_8$, on a manifold of topology $\R^8$, and to a
family of metrics, denoted by $\bB_8$, $\bB_8^+$ and $\bB_8^-$,
characterised by a non-trivial parameter (not merely a scale), on the
chiral spin bundle of $S^4$.  The metric $\bA_8$ and the metric
$\bB_8$, which is a particular case within the one-parameter family
$\bB_8^\pm$, are very simple in form, with metric coefficients that
can be expressed as rational functions of a suitably-chosen radial
variable.  The remainder of the $\bB_8^\pm$ family are more
complicated in form, and are expressed in terms of hypergeometric
functions.  All of these solutions found in \cite{cglpnew} have the
feature that $a_3$ tends to a constant at large distance, while the
remaining metric functions $a_1$, $a_2$ and $b$ have linear growth.
Thus we have, at large distance, $h_0\rightarrow 1$, $A_0\sim r$,
$C_0\sim r$ and $B_0^2\sim m^2$.

   For the $\bA_8$ metric, with $r\ge\ell>0$, we have \cite{cglpnew}
\bea
&&h_0 = \fft{(r+\ell)}{\sqrt{(r+3\ell)\, (r-\ell)}}\,,\quad
A_0=\ft12 \sqrt{(r+3\ell)\, (r-\ell)}\,,\nn\\
&&B_0= -\fft{\ell\, \sqrt{(r+3\ell)\,(r-\ell)}}{(r+\ell)}\,,\quad
C_0= \ft1{\sqrt2}\,\sqrt{(r^2-\ell^2)}\,.\label{a8sol}
\eea
From (\ref{A1sol}) we therefore have the leading-order behaviour
\be
A_1\sim \exp \Big[2\int \fft1{B_0}\Big] = e^{-4r/\ell}\,.
\ee
It follows from this that our assumption that $A_1$ is a small perturbation
also requires that we have $\ell>0$.  

   The other simple solution, $\bB_8$, corresponds to changing the
sign of $\ell$ in (\ref{a8sol}), and now taking $r\ge -3\ell>$
\cite{cglpnew}.  Defining the positive scale parameter 
$\td\ell\equiv -\ell$, we therefore have
\bea
&&h_0 = \fft{(r-\td\ell)}{\sqrt{(r-3\td\ell)\, (r+\td\ell)}}\,,\quad
A_0=\ft12 \sqrt{(r-3\td\ell)\, (r+\td\ell)}\,,\nn\\
&&B_0= \fft{\td\ell\, \sqrt{(r-3\td\ell)\,(r+\td\ell)}}{(r-\td\ell)}\,,\quad
C_0= \ft1{\sqrt2}\,\sqrt{(r^2-\td\ell^2)}\,,\label{b8sol}
\eea
with $\td\ell>0$.
From (\ref{A1sol}) we now find the leading-order behaviour
\be
A_1\sim e^{2\int \ft1{B_0}} = e^{4r/\td\ell}\,.
\ee
Thus in this case we find that we must have $\td\ell<0$
in order to have a decaying perturbation at
infinity.\footnote{The crucial distinction between the $\bA_8$ and
$\bB_8$ cases is that in $\bA_8$, the first-order equations imply that
$B_0$ has a minus-sign prefactor, as in (\ref{a8sol}), whereas in
the $\bB_8$ case the first-order equations imply that $B_0$ has a
plus-sign prefactor, as in (\ref{b8sol}).}  However, as we saw above,
with this sign the unperturbed solution is really just the $\bA_8$
metric again.

   An analogous calculation taking the whole non-trivial one-parameter
family of metrics $\bB_8^\pm$ as the zeroth-order starting points
would be more complicated.  However, by continuity we can argue that
the perturbations around these metrics would also require analogous
asymptotic behaviour, with the same signs for the coefficients of the
leading-order radial dependence as is implied by having $\td\ell$
negative in the particular case of the $\bB_8$ metric (\ref{b8sol}).
We shall, for convenience, refer to this analogous (``trivial'') scale
parameter in the general class of $\bB_8^\pm$ metrics as being the 
``$\td\ell$-parameter.''  Since there is no
family of ``$\bA_8^\pm$ metrics'' with a non-trivial parameter, we
conclude that, viewed as perturbations around the large-distance limit
of the general $\bB_8^\pm$ metrics, the $\bC_8$ metrics correspond to
``negative-$\td\ell$'' $\bB_8^\pm$ metrics that would be singular at
short distance.  The situation in these cases is somewhat analogous to
the Atiyah-Hitchin metric, in that for the perturbation to be
exponentially decaying at large distance, the sign of the scale
parameter $\td\ell$ must be {\it opposite} to the sign that is needed
for regular short-distance behaviour in the unperturbed metric.

   The perturbative discussion above does not {\it per se} allow us to
investigate directly how the perturbed solutions will extrapolate down
to short distance.  Rather, this information is contained in the
details of the $\bC_8$ solutions themselves.  The $\bC_8$ metrics
behave asymptotically like $\bA_8$ with positive $\ell$, together with
exponentially-small corrections.  They also behave asymptotically like
$\bB_8$ (or the $\bB_8^\pm$ generalisations) with negative $\td\ell$,
again with exponentially-small corrections.  Thus the $\bC_8$ metrics
can in a sense be thought of as ``resolutions'' of
(exponentially-corrected) ``negative-$\td\ell$'' $\bB_8^\pm$ metrics,
in which the singularity that one would encounter at short distance in
such a $\bB_8^\pm$ metric is smoothed out into a regular collapse to a
$\CP^3$ bolt.  Instead, the $\bC_8$ metrics can also be thought of as
a class of alternative smooth inward extrapolations from the
(exponentially-corrected) asymptotic behaviour of the $\bA_8$ metric
with positive $\ell$.  In this viewpoint the $\bC_8$ metrics are not
``resolutions,'' since the $\bA_8$ metric has positive $\ell$ and is
already itself regular, but they do provide an alternative
smooth short-distance behaviour, with a $\CP^3$ bolt instead of a NUT.
An exception to the picture of $\bC_8$ as a resolution of a
``negative-$\td\ell$'' $\bB^\pm$ metric therefore occurs if we
consider the special case of the simple $\bB_8$ metric (\ref{b8sol}),
since in this particular case the negative-$\td\ell$ $\bB_8$ metric
itself happens to be perfectly regular, being nothing but the $\bA_8$
metric.

   It is not entirely clear how one should interpret the above results
in terms of brane solutions.  In appendix \ref{ahasymp}, it was shown
how the M-theory solution of a product of seven-dimensional Minkowksi
spacetime and the Atiyah-Hitchin metric acquires an interpretation as
a negative-mass D6-brane orientifold in ten dimensions.  However, when
one is considering metrics that are asymptotically conical or locally
conical, as opposed to asymptotically flat metrics such as Taub-NUT or
Atiyah-Hitchin, the meaning of ``mass'' becomes less clear.
Typically, in the AC or ALC cases, the metric coefficients approach
constants at infinity with lesser inverse powers of the proper
distance $\rho$ than would be the case for asymptotically flat
metrics.  Asymptotically flat $d$-dimensional metrics approach
flatness like $\rho^{-d+3}$, and the ``mass'' is essentially a measure
of the coefficient of such an asymptotic term relative to the fiducial
flat metric.  It is not clear what the analogous zero-mass fiducial
metric should be when one is considering an AC or ALC metric, and
therefore it is unclear whether any coefficient in the asymptotic form
of the metric characterises the ``mass.''

   Although the understanding of mass is problematical in the present
case, we can, nevertheless, still discuss an orientifold
interpretation.  This forms the topic of section \ref{orientsec} below.


\subsection{Orientifolds}\label{orientsec}

\subsubsection{Calabi metric on line bundle over $\CP^3$}

   Before discussing the more complicated $\bC_8$ metrics with their 
Atiyah-Hitchin style ``slumping,''  
it is helpful to make contact with a previously
understood  and simpler situation, namely the metric on the line bundle over
$\CP^3$, first described in \cite{Calabi}, which, as discussed in 
\cite{g2spin7}, is the $\lambda^2=4$ limit of the $\bC_8$ metrics. 
The global structure was
discussed in detail in \cite{GibbonsFreedman}. The metric is Ricci
flat and K\"ahler, with  holonomy $SU(4) \equiv$ Spin(6) $\subset$ Spin(7),
and it is ALE, being asymptotic to $\bC^4/\bZ_4$. The
isometry group is $U(4)$, which acts holomorphically. Indeed, in 
\cite{GibbonsFreedman} a set of complex coordinates $Z^\alpha$ was
introduced, in terms of which the $U(4)$ action is linear.

    This $SU(4)$ holonomy metric is
a special solution of our equations with $a_2=-a_3=2b$ (see equations 
(10), (11) and (12) in \cite{g2spin7}).
We have (setting the arbitrary scale $\ell=1$)
\be
ds^2 =\Big( 1-{ 1\over r^8}\Big)^{-1} \, dr^2 +r^2 \, 
\Big( 1-{1 \over r^8}\Big) \, R_1^2
+ r^2 \, ({R_2}^2 + {R_3}^2 + \ft14  P_a^2)\,.
\ee
The metric
function $a_1$ vanishes at $r=1$, which is the zero section of the
line bundle. Near $r=1$ we have ( with $dt= d(r-1)\, (\sqrt 8
\sqrt{r-1})^{-1}$),
\be
ds^2\sim dt^2 + 16 t^2 \, {R_1}^2  + 
  ( {R_2}^2 + {R_3} ^2 + \ft14 {P_a}^2 )\label{near}\,.
\ee
By contrast, near infinity we have
\be
ds^2\sim dt^2 + t^2 \, (R_1^2 + R_2^2 + R_3^2 + \ft14 P_a^2 )\label{far}\,.
\ee

   The metric in round brackets in (\ref{far}) is the round metric on
$S^7$. The 1-forms $R_i$ span the $S^3\equiv SU(2)$ fibres of the
usual quaternionic Hopf fibration of $S^7$ with base $S^4$. The 1-form
$R_1$ is tangent to the fibres of the $S^1 \equiv U(1) \subset SU(2)
\equiv S^3$ complex Hopf fibration of $S^7$ with base $\CP^3$. The
appearance of the factor $16$ in the metric (\ref{near}) near the zero
section shows that the Hopf fibres must have one quarter their
standard period. Thus if we introduce Euler angles $(\tilde \theta,
\tilde \phi, \tilde \psi)$ for $SU(2)$ near the bolt, and set $R_1=
{\tilde \sigma}_3$, then the period of $\tilde \psi$ will be $\pi$.

   In the asymptotic region, the metric is manifestly asymptotically
locally flat when written in terms of the the complex coordinates
$Z^\alpha $, $ \alpha =1,2,3,4$, and the points $Z^a$ and $\sqrt{-1}\,
Z^\alpha$ must be identified. The metric may therefore be viewed as a
resolution or blow-up of the orbifold obtained by identifying flat
Euclidean space ${\Bbb E} ^8 \equiv {\Bbb E}^2 \oplus {\Bbb E}^2
\oplus {\Bbb E}^2 \oplus {\Bbb E}^2$ under a simultaneous rotation
through ninety degrees in four orthogonal two-planes. Note that the
identification does not act on the $S^4$ base of the fibration. This
is also true of the identification in the Atiyah-Hitchin case. The
identification map (a shift of $\tilde \psi$ leaving the coordinates
$\tilde \theta$ and $\tilde \phi$ on the $S^2$ base invariant) does
induce a rotation of the Cartan-Maurer 1-forms.  (This is the
identification induced by $\wtd I_3$, as discussed in appendix
\ref{ahback}.)  Thus if one seeks, as with the case of the
Atiyah-Hitchin metric, to understand the identification at the
Lie-algebra level, one must anticipate an induced action on the
$P_a$.

   As discussed in appendix \ref{ahback}, shifting $\tilde \psi$
by $\pi$ induces the action of the involution ${\wtd I}_3$ on the
Lie-algebra of $SU(2)$, namely
\be (R_1,R_2,R_3) \longrightarrow (R_1, -R_2, -R_3)\,.
\ee
However this action of ${\tilde I}_3$ alone is not an involution of
the full algebra given in (\ref{so5d}).  Inspection
reveals that one must supplement it with the following action on the
$P_a$ and $L_i$:
\be
(P_0, P_1,P_2,P_3, L_1, L_2, L_3) \longrightarrow (P_0,P_1, -P_2, -P_3,
L_1, -L_2, -L_3).  \label{extra}
\ee

\subsubsection{The $\bC_8$ metrics}

   The idea now is to use the results above to understand what happens
in the ALC rather than ALE case, where ``slumping'' takes place.
Equations (\ref{onbolt}) then apply near the bolt at $t=0$.  The
situation is somewhat more complicated than in the Atiyah-Hitchin case
discussed in the appendix, because the left-invariant 1-forms $R_i$
and $P_a$ satisfy a more complicated algebra, given in (\ref{so5d}).  
However, we can still think of the 1-forms $R_i$ as being essentially
like the standard left-invariant 1-forms $\sigma_i$ of $SU(2)$; taking
account of normalisation factors, we shall have
\be
R_i \sim -\ft12 \sigma_i + \cdots\,,
\ee
where the $\sigma_i$ are defined as in appendix \ref{ahglobal}, and
the ellipses represent the BPST Yang-Mills $SU(2)$ instanton 
connection terms that characterise the twisting
of the $S^3$ fibres over the $S^4$ base spanned by $P_a$.  

   As in appendix \ref{ahglobal}, we can make two different
``adapted'' choices for Euler angles parameterising the $S^3$ fibres,
with tilded angles near the $\CP^3$ bolt, and untilded angles at
infinity.  The vanishing of $a_1\sim 4t$ forces an identification of
the adapted Euler angle $\tilde \psi$, with period $\pi$.  Passing to
infinity, we find that $a_3$ tends to a constant. Thus if
$(\theta,\phi, \psi)$ are adapted Euler angles at infinity, the
identification that is forced is a reflection in $\psi$ and inversion
in $(\theta,\phi)$ just as in equation (\ref{i1id}) for the
Atiyah-Hitchin case.  We can think of $\psi$ as parameterising an
M-theory circle, and so the reflection of $\psi$ corresponds to
M-theory charge-conjugation, $C_{11}$.

    The inversion and additional action on the $P_a$ given by
(\ref{extra}) may be understood as follows. Reduction on the M-theory
circle leads asymptotically to a 7-manifold of cohomogeneity one with
principal orbits $\CP^3$, which we think of as an $S^2$ bundle over
$S^4$. In fact $\CP^3$ is the $\frak{ur}$-twistor space
of $S^4$. Acting on the coordinates, our involution ${\tilde I}_3$
leaves the points on the $S^4$ base fixed and acts as the antipodal
map on the $S^2$ fibres.  This ``real structure'' plays an important
r\^ole in Twistor constructions, although we shall not make use of it
in that way here.

   In the corresponding construction for the Atiyah-Hitchin metric itself, we
have, upon reduction of the negative mass Taub-NUT limit near infinity, a
3-manifold which is a cone over $S^2$. This cone is the same as ${\Bbb
R}^3$, and the antipodal map on $S^2$ induces the standard inversion (or
parity) map $P$ on ${\Bbb R}^3$. The combination of the M-theory charge
conjugation $C_{11}$ and the 3-parity $P$ amounts to inversion in four
dimensions. Since Taub-NUT is thought of as a D6-brane
\cite{Townsend}, one therefore thinks of Atiyah-Hitchin as an
orientifold plane.

   In the case of the Spin(7) manifold, we need to take the product
with ${\Bbb E}^{2,1}$ to get an eleven-dimensional Ricci-flat
solution.  We then interpret the M-theory quotient as a type IIA
solution with a D6-brane wrapped over the Cayley 4-sphere.  There are
three transverse directions; the radius $r$, and the 2-sphere with
coordinates $\theta$ and $\phi$.  The six spatial world-volume directions
consist of the 4-sphere and the two flat spatial coordinates in the
${\Bbb E}^{2,1}$ factor. The identification we are obliged to make is
thus clearly an inversion in the transverse directions.  The
orientifold interpretation therefore goes through in a completely
parallel way, and we now have an orientifold plane with a D6-brane
wrapped around the $S^4$.

\section{New $G_2$ metrics $\bC_7$ with $S^3\times S^3$ principal orbits}

    We now turn to an analogous discussion of a general class of
solutions for 7-metrics with $G_2$ holonomy.  Specifically, we shall
consider the system of first-order equations for metrics of
cohomogeneity one with $S^3\times S^3$ principal orbits.  A rather
general ansatz involving six radial functions was considered in
\cite{cglpg2,brgogugu}, and the first-order equations for $G_2$
holonomy were derived.  The six-function metric ansatz is given by
\be
ds_7^2 = dt^2 + a_i^2\, (\sigma_i-\Sigma_i)^2 + b_i^2\, (\sigma_i +
\Sigma_i)^2\,,\label{7met2}
\ee
where $\sigma_i$ and $\Sigma_i$ are left-invariant 1-forms for two
$SU(2)$ group manifolds.  It was found that for $G_2$ holonomy,
$a_i$ and $b_i$ must satisfy the first-order equations
\bea
\dot a_1 &=& \fft{a_1^2}{4 a_3\, b_2} + \fft{a_1^2}{4 a_2\, b_3}
      - \fft{a_2}{4b_3}  -\fft{a_3}{4b_2} - \fft{b_2}{4 a_3} -
        \fft{b_3}{4a_2}\,,\nn\\
\dot a_2 &=& \fft{a_2^2}{4 a_3\, b_1} + \fft{a_2^2}{4 a_1\, b_3}
      - \fft{a_1}{4b_3}  -\fft{a_3}{4b_1} - \fft{b_1}{4 a_3} -
        \fft{b_3}{4a_1}\,,\nn\\
\dot a_3 &=& \fft{a_3^2}{4 a_2\, b_1} + \fft{a_3^2}{4 a_1\, b_2}
      - \fft{a_1}{4b_2}  -\fft{a_2}{4b_1} - \fft{b_1}{4 a_2} -
        \fft{b_2}{4a_1}\,,\nn\\
\dot b_1 &=& \fft{b_1^2}{4 a_2\, a_3} - \fft{b_1^2}{4 b_2\, b_3}
      - \fft{a_2}{4a_3}  -\fft{a_3}{4a_2} + \fft{b_2}{4 b_3} +
        \fft{b_3}{4b_2}\,,\label{sixfo}\\
\dot b_2 &=& \fft{b_2^2}{4 a_3\, a_1} - \fft{b_2^2}{4 b_3\, b_1}
      - \fft{a_1}{4a_3}  -\fft{a_3}{4a_1} + \fft{b_1}{4 b_3} +
        \fft{b_3}{4b_1}\,,\nn\\
\dot b_3 &=& \fft{b_3^2}{4 a_1\, a_2} - \fft{b_3^2}{4 b_1\, b_2}
      - \fft{a_1}{4a_2}  - \fft{a_2}{4a_1} + \fft{b_1}{4 b_2} +
        \fft{b_2}{4b_1}\,.\nn
\eea

\subsection{Analysis for $\bC_7$ metrics with six-function solutions}

   We can look for regular solutions numerically, by first
constructing solutions at short distance expanded in Taylor series,
and then using these to set initial data just outside the bolt, for
numerical integration.  Once can consider various possible collapsing
spheres, namely $S^1$, $S^2$ or $S^3$.  The case $S^3$ has been
studied previously, and the case $S^2$ gives no regular short-distance
solutions.  For a collapsing $S^1$, however, we find that there are
regular short-distance Taylor expansions.  The bolt at $t=0$ will 
then be $(S^3\times S^3)/S^1$ with the circle embedded diagonally; this
is the space $T^{1,1}$.  By analogy with the eight-dimensional $\bC_8$
metrics of Spin(7) holonomy with a collapsing circle at short
distance, we shall denote these new $G_2$ metrics by $\bC_7$. 
 
   Taking the collapsing $S^1$, without loss of generality, to be in
the $a_3$ direction, we find regular short-distance solutions of the
form
\bea
a_1 &=& q_1 - \fft{(q_1^2 -q_2^2 +q_3^2)}{8 q_1\, q_3}\, t +
O(t^2)\,,\nn\\
a_2 &=& q_1 + \fft{(q_1^2 -q_2^2 -q_3^2)}{8 q_2\, q_3}\, t +
O(t^2)\,,\nn\\
a_3 &=& -t - \fft{[(q_1^2 - q_2^2)^2 +q_3^4 
            -8 (q_1^2+ q_2^2)\, q_3^2]}{96 q_1^2\, q_2^2\, q_3^2}\,
                     t^3 + O(t^5)\,,\nn\\
b_1 &=& q_1 -\fft{(q_1^2 - q_2^2 - q_3^2)}{8 q_2\, q_3}\, t + O(t^2)\,,\nn\\
b_2 &=&  q_2 +\fft{(q_1^2 - q_2^2 +q_3^2)}{8 q_1\, q_3}\, t + O(t^2)\,,\nn\\
b_3 &=& q_3 - \fft{[(q_1^2 -q_2^2)^2 - q_3^4]}{16 q_1^2\,
q_2^2\, q_3}\, t^2 + O(t^4)\,.\label{g2taylor}
\eea
Here $(q_1,q_2,q_3)$ are free parameters in the solutions. They
characterise the metric on the $T^{1,1}$ space that forms the bolt.
One of the three parameters corresponds just to setting the overall
scale of the metric.

    We have calculated the Taylor expansions (\ref{g2taylor}) up to
tenth order in $t$, and used these in order to set initial data just
outside the $T^{1,1}$ bolt.  Numerical integration can then be used in
order to study the possibility of having solutions that remain regular
at large $t$.  The numerical analysis of the system of six equations
(\ref{sixfo}) seems to be somewhat delicate, and the results are
rather less stable than one would wish, but they seem to suggest that
if we fix, say, $q_1$ and $q_2$, then there exists a a range of values
$q_3\le K$, for some constant $K$, that gives regular solutions.  The
indications are that the metrics will be ALC, for $q_3<K$, and AC for
$q_3=K$.  For the ALC metrics it is the function $b_3$ that
stabilises to a fixed radius at large distance.  Thus at short
distance the circle spanned by $(\sigma_3-\Sigma_3)$ collapses, while
at large distance the circle spanned by $(\sigma_3+\Sigma_3)$
stabilises.  This means that whilst the ALC metrics have a $T^{1,1}$
bolt at short distance, and approach $S^1$ times a cone over $T^{1,1}$
at large distance, the $T^{1,1}$ spaces in the two regions correspond
to two different embeddings of $S^1$ in $S^3\times S^3$.  Note that
the metric coefficient $b_3^2$ remains finite and non-zero everywhere,
both at short distance and large distance.

   If we choose $q_1=q_2$, the equations (\ref{sixfo}) imply that 
we shall have $a_1=a_2$ and $b_1=b_2$ for all $t$, and in fact the
six-function system of equations can then be consistently truncated to
a four-function system.  As we describe in section \ref{fourfnsec}
below, this truncated system of equations seems to give much more
stable and reliable numerical results.  

   One can also repeat the perturbative analysis at large distance
that we applied previously to the Atiyah-Hitchin metric (in appendix
A) and to the Spin(7) metrics in section 2.  We shall look here for
solutions at large distance that are perturbations around the exact
solution found in \cite{brgogugu}, for which we shall make the ansatz
\bea
&&
a_1=\ft{\sqrt3}4 \sqrt{(r-\ell)(r+3\ell)} +x\,,\quad
a_2=\ft{\sqrt3}4 \sqrt{(r-\ell)(r+3\ell)} -x\,,\nn\\
&&b_1=-\ft{\sqrt3}4\, \sqrt{(r+\ell)(r-3\ell)} + y\,,\qquad
b_2=-\ft{\sqrt3}4\, \sqrt{(r+\ell)(r-3\ell)} - y\,,\\
&&a_3=-\ft12 r\,,\qquad
b_3=\fft{\ell\, \sqrt{r^2-9\ell^2}}{\sqrt{r^2-\ell^2}}\,,
\eea
where $dr = -\ft{2b_3}{3\ell}\, dt$, and $x$ and $y$ are assumed to be
small in comparison to the leading-order terms at infinity.  At the
linearised level at infinity, we find that the functions $x$ and $y$ are
approximately given by
\be
x\sim \fft{1}{r^2}\, \Big(c_1\, e^{\ft{3r}{2\ell}} + c_2\, e^{
-\ft{3r}{2\ell}}\Big)\,,\qquad
y\sim \fft{6c_1}{\ell\, r}\, e^{\ft{3r}{2\ell}} -
\fft{c_2}{r^2}\, e^{-\ft{3r}{2\ell}}\,.
\ee
Thus unlike the previous Spin(7) case, we cannot derive the sign of
the mass at infinity, since with an appropriate choice of the
constants $c_1$ and $c_2$, a small perturbation can give either sign.

\subsection{Analysis for $\bC_7$ metrics for the
four-function truncation}\label{fourfnsec}

\subsubsection{Numerical results}

   The six-function set of equations (\ref{sixfo}) can be consistently
truncated to a four-function system, by setting corresponding pairs of
the $a_i$ and $b_i$ equal.  Choosing, for example, $a_2=a_1$ and
$b_2=b_1$, we obtain
\bea
\dot a_1 &=& \fft{a_1^2}{4a_3\, b_1} - \fft{a_3}{4b_1} -
\fft{b_1}{4a_3} - \fft{b_3}{4a_1}\,,\qquad
\dot a_3 = \fft{a_3^2}{2a_1\, b_1} - \fft{a_1}{2b_1} - \fft{b_1}{2a_1}
\,,\nn\\
\dot b_1 &=& \fft{b_1^2}{4a_1\, a_3} -\fft{a_1}{4a_3} -
\fft{a_3}{4a_1} + \fft{b_3}{4b_1}\,,\qquad
\dot b_3 = \fft{b_3^2}{4a_1^2} - \fft{b_3^2}{4b_1^2}\,.
\eea
This system is easier to analyse, and for the purposes of a numerical
analysis, we find that we get much more stable and reliable results.

     We shall again look for solutions where there is a
collapsing $S^1$ at short distance.  The Taylor expansions are in
fact those following from (\ref{g2taylor}) by setting $q_1=q_2$.  Without
loss of generality we shall make a scale choice, and set $q_1=q_2=1$, 
and take $q_3=q$, giving 
\bea
a_1 &=& 1 - \ft18 q\, t + \ft1{128}(16-3q^2)\, t^2 + \cdots
\,,\nn\\
a_3 &=& -t -\ft{1}{96} (q^2-1)\, t^3 + \cdots
\,,\nn\\
b_1 &=& 1 + \ft18 q\, t + \ft{1}{128}(16-3q^2)\, t^2 + \cdots
\,,\nn\\
b_3 &=& q + \ft1{16} q^3\, t^2 + \cdots\,.
\eea
Numerical analysis indicates that the solution is regular if $|q|\le
q_0=0.917181\cdots$, with $|q|=q_0$ giving an AC solution, and
$|q|<q_0$ giving ALC solutions in which $b_3$ becomes constant at
infinity.\footnote{The metric on the $T^{1,1}$ bolt at $t=0$ would itself
be Einstein if $q=2/\sqrt3\sim 1.154701\cdots$, and so all of the
non-singular seven-dimensional metrics correspond to situations where
the $T^{1,1}$ bolt is squashed along its $U(1)$ fibres, relative to the
length needed for the Einstein metric.}
Again we see that at short distance the circle spanned by
$(\sigma_3-\Sigma_3)$ collapses, while at large distance the circle
spanned by $(\sigma_3+\Sigma_3)$ stabilises.  Thus here too we see
that the ALC metrics have a $T^{1,1}$ bolt at short distance, and
approach $S^1$ times a cone over $T^{1,1}$ at large distance, but the
$T^{1,1}$ spaces in the two regions correspond to two different
embeddings of $S^1$ in $S^3\times S^3$.  Again, we have the feature that
the metric coefficient $b_3^2$ is finite and non-zero everywhere, including
short distances and large distances.

\subsubsection{Global structure of $\bC_7$ metrics}\label{glob7}

     To understand the effect of this new type of ``slump'' in the
$G_2$ manifolds, it is useful to express the $SU(2)$ left-invariant
1-forms in terms of Euler angles:
\bea
&&\sigma_1 = \cos\psi_1\, d\theta_1 +
\sin\psi_1\,\sin\theta_1\, d\phi_1\,,\qquad
\Sigma_1 = \cos\psi_2\, d\theta_2 +
\sin\psi_2\,\sin\theta_2\, d\phi_2\,,\nn\\
&&
\sigma_2 = -\sin\psi_1\, d\theta_1 +
\cos\psi_1\,\sin\theta_1\, d\phi_1\,,\qquad
\Sigma_2 = -\sin\psi_2\, d\theta_2 +
\cos\psi_2\,\sin\theta_2\, d\phi_2\,,\nn\\
&&
\sigma_3 = d\psi_1 + \cos\theta_1\, d\phi_1\,,\qquad
\Sigma_3 = d\psi_2 + \cos\theta_2\, d\phi_2\,.
\eea
Making the following redefinitions,
\be
\psi=\psi_1 + \psi_2\,,\qquad
\td \psi = \psi_1-\psi_2\,,
\ee
we have
\bea
ds_7^2
&=& dt^2 + (a_1^2 + b_1^2)\, \Big(d\theta_1^2 +d\theta_2^2 +
\sin^2\theta_1\, d\phi_1^2 + \sin^2\theta_2\, d\phi_2^2\Big)\nn\\
&&-2(a_1^2 -b_1^2)\Big(\sin\td\psi\, (\sin\theta_1\, d\theta_2\,d\phi_1 -
\sin\theta_2\, d\theta_1\, d\phi_2) \nn\\
&&\qquad\qquad\qquad+
\cos\td\psi\, (d\theta_1\, d\theta_2 + \sin\theta_1\, \sin\theta_2\,
d\phi_1\, d\phi_2)\Big)\nn\\
&& +a_3^2\, (d\td\psi + \cos\theta_1\, d\phi_1 -\cos\theta_2\, d\phi_2)^2
+ b_3^2\, (d\psi + \cos\theta_1\, d\phi_1 +\cos\theta_2\, d\phi_2)^2\,.
\label{eulermetric}
\eea
At small distance, the geometry is an $\R^2$ bundle over a 
squashed $T^{1,1}/\Z_N$, with $\psi$, the fibre coordinate, having any
period of the form $4\pi/N$.  Regularity of the collapsing
circles requires that $\td\psi$ must have period of $2\pi$.  At large
distance, since the maximum allowed period for $\td\psi$, which forms
the fibre coordinate in the $T^{1,1}$ there, would have been $4\pi$,
this means that we  have a $T^{1,1}/\Z_2$ at large distance.

    It is worth remarking that from (\ref{eulermetric}) we can see
that $K\equiv \del/\del\psi$ is an exact Killing vector everywhere.
Furthermore, as we already noted, the metric coefficient $b_3^2$ is
finite and non-vanishing everywhere, which means that the length of 
$K$ is everywhere finite and non-zero.  As far as we are aware, there
is no other known example of a Ricci-flat metric with such a $U(1)$ 
isometry, and in which the circle is not a metric product.  In fact
there are necessary conditions that must be satisfied by any space
that admits such a property, which we can verify are indeed satisfied
by $\bC_7$.  Firstly, the Euler number must be zero.  Since $\bC_7$ is 
an $\R^2$ bundle over $T^{1,1}$, and $T^{1,1}$ is itself topologically
$S^2\times S^3$, it follows, since $S^3$ has no relevant non-trivial
topology, that $\bC_7$ is topologically the product of $S^3$ with an
$\R^2$ bundle over $S^2$.  Since $S^3$ has zero Euler number, it
follows that $\bC_7$ has also.  A second necessary condition is that
the $U(1)$ Killing vector must not be hypersurface orthogonal.  In
other words, viewed as a 1-form, we must have $K\wedge dK\ne0$.  It is
easily verified that indeed $K\wedge dK$ is non-zero in our case.

\subsubsection{Reduction to a type IIA solution}

   Let us consider an eleven-dimensional ``vacuum'' comprising the direct
product of four-dimensional Minkowski spacetime and a $\bC_7$ manifold
of $G_2$ holonomy.   We observed in section \ref{glob7} that the
length of the Killing vector $K=\del/\del\psi$ is everywhere finite
and non-zero, and this means that we can perform a Kaluza-Klein reduction on
the circle parameterised by $\psi$ that will be completely
non-singular.

   By reducing on the coordinate $\psi$ we then
obtain a solution of the type IIA theory, with
metric
\be
ds_{10}^2= e^{-\fft16\phi}\, dx^\mu\,
dx_\mu + e^{\fft43\phi}\, ds_6^2\,.\label{d10metric}
\ee
The
period of $\psi$ may be taken to be $4\pi/N$.  The 
string coupling constant is given by
\be
g_{\rm str} = e^{\phi} = \Big(\fft{b_3}{N}\Big)^{3/2}\,,
\ee
and so it is finite and non-vanishing
everywhere.  It follows that if $N$ is large, the solution is a good 
approximation at the string theory level, both at large and small 
distances.  The geometry of the reduced 6-metric is $\R^2\times
S^2\times S^2$ at small distance, and the cone over $T^{1,1}/Z_2$ at large
distance.  The R-R 2-form in the type IIA solution carries magnetic
flux threading the two $S^2$ cycles. 

    The circle parameterised by $\psi$ in this $G_2$ manifold also
provides a stable $S^1$ that can be used for performing a T-duality
transformation between the type IIA and type IIB theories, 
without encountering any singularity.

    To close this discussion of the dimensionally-reduced metric, 
we remark that, as discussed in \cite{cglpg2}, the $\sigma_i$ and
$\Sigma_i$ left-invariant $SU(2)$ 1-forms can be embedded within the
left-invariant 1-forms $L_{AB}$ of $SO(4)$ as
\bea
&&\sigma_1 = L_{42} + L_{31}\,,\qquad
  \sigma_2 = L_{23} + L_{41}\,,\qquad
  \sigma_3 = L_{34} + L_{21}\,,\nn\\
&&
\Sigma_1 = L_{42} - L_{31}\,,\qquad   \Sigma_2 = L_{23} - L_{41}
\,,\qquad \Sigma_3 = L_{34} - L_{21}\,,
\eea
where $dL_{AB}=L_{AC}\wedge L_{CB}$.  In terms of this basis, the
six-dimensional metric in (\ref{d10metric}) takes the form
\be
ds_6^2 = dt^2 + \td a^2\, L_{1i}^2 + \td b^2\, L_{2i}^2 + \td c^2
L_{12}^2\,,\label{stenform}
\ee
where $\td a$, $\td b$ and $\td c$ are functions of a radial variable
$t$, and $i=3,4$.  The metric (\ref{stenform}) is written precisely in
the form used in \cite{cglpsten} for writing the the Stenzel metric on
the cotangent bundle of $S^3$ (\ie the deformation of the conifold,
first constructed in \cite{candel}).  Thus the metric (\ref{d10metric})  
can be recognised as a generalisation of the metric on the product of
(Minkowski)$_4\times$(Stenzel)$_6$, in which the dilaton and the R-R
2-form field strength of the type IIA theory are also excited.

\section{ALG gravitational instantons and periodic monopoles}

\subsection{ALG solutions}

   In a recent paper, Cherkis and Kapustin \cite{Cherkis} have
discussed  "ALG" gravitational instantons. These are
asymptotically locally invariant under a tri-holomorphic $T^2$
action and in some respects resemble the Atiyah-Hitchin metric. The reason for
their choice of  the name is presumably that G follows F
in the alphabet.  Exact expressions for these metrics are not known, 
but {\sl asymptotically}, an ALG  metric may be cast in the form
\be 
ds^2=  { 1 \over \tau_2} \, |dy_1 + \tau\,  dy_2 |^2 + \tau _2 \, dz
\, d{\overline z}\,, \label{alg} 
\ee 
where $y_1$ and $y_2$ are real, and $\tau=\tau(z)=t_1 +\im\, \tau _2$
is a holomorphic or anti-holomorphic 
function of $z$. Thus $\tau_2$ is a harmonic
function, and
the metric (\ref{alg}) is of the standard adapted form with respect to
the tri-holomorphic Killing vector field $\partial /\partial
y_1$. The Killing vector field $\partial/ \partial y_2$ is
also tri-holomorphic, and  passing to coordinates adapted to it
involves a Legendre transformation, which may be regarded as an
example of mirror symmetry. The two coordinates $y_1$ and $y_2$
are periodically identified, and so locally the metric (\ref{alg})
admits an elliptic fibration, \ie a fibration by tori labelled by
the complex coordinate $z$, and  carrying a unimodular metric
parameterised by $\tau(z)$.

   It should be emphasised that neither the asymptotic form
(\ref{alg}) nor the tri-holomorphic $T^2$ isometry group remains valid
in the interior of the manifold.  In fact in general, the exact
interior metric will have no Killing vectors at all. In special cases,
or at special points in the modulus space, it may admit
a non-tri-holomorphic circle action. In what follows we shall use
$\theta$ for the angle conjugate to this Killing field, which in
adapted coordinates takes the form $\partial / \partial \theta$.

   One of the metrics considered in \cite{Cherkis} is the relative
modulus space for two BPS Yang-Mills monopoles on ${\Bbb E}^2 \times
S^1$ with the product metric, the circle having radius $L$. In the
limit that $L$ goes to infinity it approaches the Atiyah-Hitchin
metric, which is the relative modulus space of two BPS Yang-Mill
monopoles on ${\Bbb E}^3$. Because the system is invariant under the
Euclidean group acting on ${\Bbb E} ^2$, one expects the relative
modulus space to be invariant under the action of $SO(2) \subset
E(2)$, which has the effect of rotating the monopoles about their
centre of mass. In the limit that $L$ tends to infinity, the symmetry
is enhanced to $SO(3) \supset SO(2)$. The putative $SO(2)$ action
would not be tri-holomorphic and, as we shall see in more detail
later, this would have the consequence that the metric could be written in
terms of a solution of the $su(\infty)$ Toda equation.

   The other metrics considered in \cite{Cherkis} have $n$ additional
Dirac-type monopoles present, with $n=1,2,3,4$. They show, using
the techniques of \cite{gibman}, that the appropriate
choice for $\tau$ is\footnote{This formula is taken from the revised version
of \cite{Cherkis}.   In the earlier version of this paper, we used the
formula given in version I of \cite{Cherkis}.  The correction does not
affect our results, since the behaviour at large $|z|$ is unchanged.}
\be 
\tau= \im\, \Big( {r_{\rm ren} -ib \over 2}
+ { 1\over \pi}\, \log( {\overline z}) - { 1\over 8 \pi }\, 
\sum_{j=1}^n \log ({\overline m}_j^2 -\ft14 {\overline z}^2) \Big)\,. 
\label{potxxx}
\ee
The asymptotic metric is complete as $|z|$ goes to infinity. The
volume growth is quadratic: the volume inside a region of proper
radius $r$ grows as $r^2$. If $n\ne 4$, the circle in the $y_1$
direction collapses in size while the length of the circle in the
$y_2$ direction blows up. If $n=4$, the metric is (up to
identifications) asymptotic to the flat metric on ${\Bbb C} \times
T^2$. If $m_i=0$ for $i=1,2,3,4$, the asymptotic metric is exactly
flat. In general the metric has a non-trivial ``monodromy,'' meaning
that traversing a large circle in the base induces an $SL(2, {\Bbb
Z})$ transformation of the torus. From a Kaluza-Klein perspective, the
solution behaves like a cosmic string carrying Kaluza-Klein (or in the
string context, Ramond-Ramond) flux, associated to the two asymptotic
$U(1)$ isometries. This monodromy phenomenon is therefore related to
the work of \cite{Onemli-Tekin} on ``Kaluza-Klein vortices.''  In the
case $n=4$, the fluxes vanish and hence there is no monodromy.

\subsection{Monodromy and the Heisenberg group}

   The metric (\ref{alg}) with $n=0$ admits a mono-holomorphic axial
symmetry. Together with the two tri-holomorphic symmetries we get
a three-dimensional isometry group acting on three-dimensional
orbits. In other words the metric is of cohomogeneity one. In
fact the metric is of Bianchi type II, \ie it is invariant
under the action of the Heisenberg group. Related metrics occur as
supersymmetric domain walls \cite{Stelle}.  However, by contrast
with the domain-wall  metrics, for which the Heisenberg action is
tri-holomorphic, in the present case the action is
mono-holomorphic. Roughly speaking, the distinction between the
two cases is whether the group action commutes with supersymmetry.
Another way of expressing the difference is that if the curvature
is self-dual, and $K$ is a Killing vector with $k=K_a\, dx^a$ the
associated 1-form,  then it is tri-holomorphic if  $dk$ is
self-dual, or mono-holomorphic if  $dk$ is anti-self-dual. If $
n\ne 0$ the metric (\ref{alg}) is not exactly
axisymmetric, unless $m_i=0$, but it becomes so near 
infinity. The limiting metric
therefore also admits a Heisenberg action. To see this explicitly,
note that at large distances (or exactly, if $n=0$) we
may take (after convenient rescalings and choices of inessential
constants) 
\be
\tau_1= ( 1-{n \over 4})\, \theta\,,\qquad
\tau_2 =  (1-{n \over 4} )\, \log r\,. \label{tauchoice}
\ee

   Let
\be 
\sigma_1= d y_2,  \qquad \sigma _2 =d \theta,  \qquad \sigma_3 =
 d y_1+  (1-{n \over 4} ) \,\theta \, dy_2\,.
\ee 
One has 
\be 
d\sigma_1=d\sigma_2=0, \qquad d\sigma_3= -( 1-{n
\over 4} ) \, \sigma_1 \wedge \sigma_2 \,.\label{h123} 
\ee 
Thus $(\sigma_1,
\sigma_2 ,\sigma_3)$ are left-invariant 1-forms on the
Heisenberg group, and the metric may be cast in the standard
triaxial form 
\be 
ds^2 = a^2\,  b^2\,  c^2\,  d \eta^2 +  a^2\,  \sigma_1^2 
+b^2 \, \sigma_2 ^2 + c^2 \,\sigma_3 ^2\,,  \label{Lorenz1} 
\ee 
with
\be 
r=e^\eta,  \qquad a^2=c^{-2} =  \tau_2, \qquad b^2 =r^2\, 
\tau_2.  \label{Lorenz2} 
\ee

   The Killing vectors are
\be
{\bf e}_1= \partial _{y_1} \,\qquad  {\bf e}_2 =\partial_{y_2}  \,
\qquad {\bf e}_3= \partial_\theta -(1-{n \over 4} )\, y_2\, \partial_{y_1} .
\ee
The only non-vanishing  bracket is 
\be
[{\bf e}_2, {\bf e}_3] = -(1-{n \over 4})\, {\bf e}_1.
\ee
One may check that the  metric (\ref{Lorenz1}) with
(\ref{Lorenz2})  coincides with the triaxial self-dual Bianchi II
metric (see equation (24b) of the first reference in \cite{Lorenz}, after
setting $\lambda_2=1$). The tri-holomorphic case (which has
$a^2=c^2$) is obtained by instead setting $\lambda_2=0$. This admits an
extra $U(1)$ isometry.  These two cases are analogous to the
Atiyah-Hitchin metric, which is triaxial (being invariant merely
under $SO(3)$), and the Eguchi-Hanson metric, with its isometry group
$U(2)$. The latter is biaxial and admits an additional $U(1)$

   The first-order equations implying $SU(2)$ holonomy can be derived from
a superpotential.  For the sake of completeness, we shall present the
superpotentials for the both the Bianchi IX system, when $d\sigma_i = 
-\ft12 \ep_{ijk}\, \sigma_j\wedge \sigma_k$, and the Heisenberg or Bianchi II 
case, which can be taken to have $d\sigma_1=d\sigma_2=0$, $d\sigma_3=
-\sigma_1\wedge \sigma_2$ (a simple rescaling of (\ref{h123})).  The 
conditions for Ricci-flatness for the metrics (\ref{Lorenz1}) can be
derived from the Lagrangian $L=T-V$ where both for Bianchi IX and 
Bianchi II, the kinetic terms are given by
\be
T =\ft12 g_{ij}\, \fft{d\a^i}{d\eta}\, \fft{d\a^j}{d\eta}\,,\qquad
g_{ij}= \pmatrix{0&1&1\cr 1&0&1\cr 1&1&0}\,,
\ee
where $(a,b,c)=(e^{\a^1},e^{\a^2},a^{\a^3})$.  The potentials are
\bea
\hbox{Bianchi IX}:&& V= 
  \ft14(a^4+b^4+c^4-2 b^2\, c^2 -2 c^2\, a^2 -2 a^2\, b^2) \,,\nn\\
\hbox{Bianchi II}:&& V= \ft14 c^4\,.
\eea
These potentials can be derived from superpotentials, $V=-\ft12 g^{ij}\, 
\del W/\del\a^i\, \del W/\del\a^j$, with
\bea
\hbox{Bianchi IX (1)}:&& W= \ft12 (a^2+b^2+c^2)\,,\nn\\
\hbox{Bianchi IX (2)}:&& 
    W= \ft12 (a^2+b^2+c^2- 2b\, c- 2c\, a-2 a\, b))\,,\nn\\
\hbox{Bianchi II (1)}:&& W = \ft12 c^2 + k\, a\, c\,,\nn\\
\hbox{Bianchi II (2)}:&& W = \ft12 c^2 + k\, b\, c\,,
\eea
where, for the Bianchi II cases, $k$ is an arbitrary constant.  Note 
that only for $k=0$ can the Bianchi II superpotential be obtained as a 
scaling limit of the Bianchi IX superpotentials (which both yield the
same $k=0$ limit).

   The corresponding first-order equations, re-expressed in terms of
the proper-distance coordinate $t$, for which $dt=a\, b\, c\, d\eta$,
are
\bea
\hbox{Bianchi IX (1)}:&& \dot a = \fft{b^2+c^2-a^2}{2b\, c}\,,\qquad 
\hbox{and cyclic}\,,\nn\\
\hbox{Bianchi IX (2)}:&& \dot a = \fft{(b-c)^2-a^2}{2b\, c}\,,\qquad 
\hbox{and cyclic}\,,\nn\\
\hbox{Bianchi II (1)}:&& \dot a = \fft{c}{2b}\,,\quad 
      \dot b= \fft{c}{2a} + k\,,\quad \dot c = -\fft{c^2}{2a\, b}\,,\nn\\
\hbox{Bianchi II (2)}:&& \dot a = \fft{c}{2b} + k\,,\quad 
      \dot b= \fft{c}{2a}\,,\quad \dot c = -\fft{c^2}{2a\, b}\,,
\eea
Case (1) for Bianchi IX is the system of equations whose only complete
non-singular solution is the Eguchi-Hanson metric, for which two of 
the three metric functions are equal; however, a general triaxial
solution, albeit singular, also exists \cite{begipapo}.  Case (2) for 
Bianchi IX has the Atiyah-Hitchin metrics as its general regular solution 
\cite{atihit}, with Taub-NUT as a special regular solution if any two of the
metric functions are set equal.  

   For Bianchi II (or Heisenberg), the two cases are equivalent,
modulo a relabelling of $a$ and $b$.  Up to an overall scale, all 
non-vanishing choices for the constant $k$ are equivalent.  
If we take $k=0$, the solution has the domain-wall form
\be
ds_4^2 = y\, dy^2 + y\, (\sigma_1^2+\sigma_2^2) + \fft1{y}\, \sigma_3^2
\ee
that was discussed in \cite{lalupo,gibryc,Stelle}.  If, on the other
hand, we take $k=1$, the solution has the form
\be
ds_4^2 = a_0^2\, b_0^2\, y\, e^{2 a_0\, y}\, dy^2 
  + 2a_0^2\, y\, \sigma_1^2 + 
  2 b_0^2\, y\, e^{2a_0\, y}\, \sigma_2^2 + \fft{2}{y}\, \sigma_3^2\,.
\ee
If we set $y=2\log r$, and take the constants to be $a_0=b_0=\ft12$, we get 
the $n=0$ metric of Cherkis and Kapustin, 
\be
ds^2 = \log r\, dr^2 + \log r\, \sigma_1^2 + r^2\, \log r\, \sigma_2^2 
     +\fft1{\log r}\, \sigma_3^2\,.\label{n0met}
\ee

    Note that the rotational invariance is not manifest because the
coordinate $\theta$ appears explicitly in the metric. This may be
avoided by changing to a new variable 
\be 
y_1 \longrightarrow y_1 +
(1- { n \over 4}) \, \theta \, y_2\,.\label{legendre} 
\een
Using this new coordinate, one may put the metric in canonical form
with respect to the non-tri-holomorphic Killing field $\partial
 / \partial \theta$ , and extract the relevant solution of the
Toda equation.

   The  monodromy phenomenon rests on the fact that, with suitable
identifications, one may think of the Heisenberg group as a $T^2$
bundle over $S^1$. Traversing a closed loop  in the base space
leads to an $SL(2, {\Bbb Z})$ action on the $T^2$ fibres, whose
coordinates are $(y_1, y_2)$. In the case $n=0$, going half way
around the  circle  corresponds to interchanging two identical
monopoles, and so one must identify $(z,y)$ with $(-z,-y)$.

\subsection{Masses} 

   It is important to  note that the strength of
the logarithmic potential associated with the monopole is $-4$ times
the strength of the logarithmic potential associated with each of
the Dirac singularities. A similar factor of $-4$ arises in the
case of the asymptotic metric of the $D_n$ ALF gravitational
instantons obtained by Sen \cite{Sen}. These have an asymptotic
tri-holomorphic circle action,  with associated harmonic function on
${\Bbb E} ^3$ given by
\be 
1 - { 16  M \over |{\bf x}|  } + \sum _1^n
\bigl ( { 4  M \over |{\bf x} -{\bf x}_j  | }+ { 4 M \over | {\bf
x} + {\bf x} _j | } \bigr )\,. \label{Sen} 
\ee
The first non-constant term is needed to get the asymptotic Atiyah-Hitchin
metric (\ie Taub-NUT with negative mass). In stringy language this is
the orientifold 6-plane with negative tension. The remaining terms
represent D6-branes with positive tension, with the ratio $(-4)$ of
charges being derived from string theory. The form of the
contribution of the D6-branes is that needed to make the $CP$
identification. Note that in the case of the $D_2$ metric we have
zero mass. This case appears to coincide with the Page-Hitchin
metric \cite{Page1, Hitchin1,Kobayashi1}. The total Kaluza-Klein
monopole moment vanishes, and the boundary is an identified product
$(S^1 \times S^2)/ \Gamma$.

   A similar interpretation follows for the asymptotic metrics
obtained in \cite{Cherkis}.  Note that the last term in (\ref{potxxx})
is invariant under $z\longrightarrow -z$, just as (\ref{Sen}) is
invariant under ${\bf x}\longrightarrow {\bf x}$.  It is natural to
regard them as corresponding to orientifolds wrapped over a 2-torus,
together with $n$ wrapped D6-branes.

   As we remarked earlier, there is in general no reason to suppose
that an ALG metric admits any Killing vectors at all, except in the
asymptotic large-distance limit. Certainly there can be no
tri-holomorphic isometries. As noted above, there is an $SO(2)$ action
in the case $n=0$, but this may not be manifest from the asymptotic
form of the metric because of the phenomenon of monodromy. The absence
of tri-holomorphic isometries means that the Kaluza-Klein or
Ramond-Ramond electric charges associated with the two asymptotic
$U(1)$ isometries will not be exactly conserved.  Processes in the
core will lead to their violation (see \cite{schwarz}).

\subsubsection{Olber-Seeliger paradox and negative masses}

   In this section we point out that approximate constructions using
multi-centre metrics \cite{hawk,gibhaw} will typically involve
negative mass-points.  Sufficiently close to the negative mass points,
the metric signature will become negative definite. This, of course,
signals a breakdown of the multi-centre approximation, while the exact
solution that the multi-centre metric approximates will be perfectly
regular. However, if the multi-centre approximation is good at large
distances, the long-range fields of the negative mass-points may show
up there.

   To see why negative mass-points are typically required, recall that
constructions using blow ups and Eguchi-Hanson metrics may be
summarised as follows: One considers $({\Bbb C}^2 \equiv(z,y )/
\Lambda)/\Gamma$ and blows up the singularities, where $\Lambda$ is a
lattice (a discrete abelian group of translations) and $\Gamma$ is an
involution which acts as before. At one extreme, if $\Lambda$ has rank
zero we get the Eguchi-Hanson metric itself. At the other extreme, if
$\Lambda$ has maximal rank, \ie rank 4, we get the Kummer construction
\cite{Gibbons-Pope,Page2}. As intermediate cases, if $\Lambda$ has
rank 2 we get the ${\overline D}_4$ metrics, whilst if $\Lambda$ has
rank 1 we get Page's periodic but non-stationary instanton
\cite{Page1,Hitchin1,Kobayashi2}. If $\Lambda$ has rank 3 we perhaps
get something like the quasi-periodic gravitational instantons of
\cite{Nergiz-Saclioglu}.

   If one uses the harmonic function ansatz, one is likely to run into
the problem that if all the terms are taken to be positive, then the
associated expression for the potential is that of a periodic array of
charges all of the same sign and the same magnitude, and this sum will
not converge. This is essentially the gravitational version of Olber's
(or more strictly Halley's) paradox in cosmology, and in the
gravitational context is usually ascribed to Seeliger. One way of
circumventing it is to use a hierarchical distribution along the lines
of \cite{Anderson-Kronheimer-LeBrun}. Another way is to introduce
negative, as well as positive, masses. It is easily seen from the
expressions in \cite{Nergiz-Saclioglu} and \cite{Ooguri-Vafa} that
this is indeed how they arrange to get convergent periodic potentials.

\subsection{Cosmic string solutions}

    In this sub-section we shall relate the  ALG metrics to Stringy
Cosmic Strings \cite{Greene}, Dirichlet Instanton corrections
\cite{Ooguri-Vafa,Gross-Wilson}, the seven-brane of type
IIB theory \cite{gigrpe}, periodic gravitational instantons
\cite{Nergiz-Saclioglu}, and work on Kaluza-Klein vortices
\cite{Onemli-Tekin}.

  From a type IIB perspective, we write the ten-dimensional metric
in Einstein gauge as 
\ben ds^2 = -dt^2 + (dx_9)^2
+(dx_8)^2+(dx_7)^2+(dx_6)^2+(dx_5)^2+(dx_4)^2+(dx_3)^2 + e^{ \phi}
dz \, d{\bar z}. 
\een 
The static equations arise from the
two-dimensional Euclidean Lagrangian 
\ben 
L= R-{(\partial
\tau_1)^2 +(\partial \tau_2)^2 \over 2\tau_2^2}\,, 
\een 
where
$\tau=\tau_1+ \im\, \tau_2= a+ \im\,  e^{-\Phi}$ gives a map into the
fundamental domain of the modular group $SL(2,{\Bbb Z}) \backslash
SO(2, {\Bbb R}) /SO(2)$. We may regard  the two-dimensional spatial
sections as a K\"ahler manifold, and the harmonic map equations are
thus satisfied  by the {\sl holomorphic ansatz}  $\tau= \tau(z)$. We
must also satisfy the Einstein condition. Using the formula for
the Ricci scalar of the two-dimensional metric, and the holomorphic
condition, this reduces to the {\sl linear} Poisson equation 
\ben
\partial {\bar \partial }( \phi - {\log} \tau_2 )=0.
\een

   To get the {\it fundamental string}
one interprets the axion and dilaton as coming from the NS-NS sector.
On therefore  chooses 
\ben 
\phi=\Phi \,,\qquad \tau \propto \log z\,. 
\een
In four spacetime dimensions the fundamental string is 
``super-heavy ,'' and it is not asymptotically conical at infinity.

   To get the {\it seven brane}, which does correspond to a more
conventional cosmic string,  one picks 
\ben 
j(\tau(z)) = f(z)= { p(z) \over q(z) }\,, 
\een 
where $j(\tau)$ is the elliptic modular
function and $ f(z)$ is a rational function of
degree $k$.

   The appropriate solution for the metric is 
\ben 
e^{\phi} = \tau_2\, 
\eta^2\, {\bar \eta}^2 \, \Bigl |  \prod_{i=1}^k (z-z_i) ^ { -{1 \over
12} } \Bigl |^2\,,
\een 
where $\eta (\tau)$ is the Dedekind function. Asymptotically
\ben 
e^\phi \sim (z {\bar z} )^{-{ k/12}}\,, 
\een 
and therefore
the  spatial metric is that of a cone with deficit angle 
\ben
\delta = { 4k \pi \over 24}\,. 
\een 
This may also be verified using
the equations of motion and the Gauss-Bonnet theorem.  As a result,
one  can have up to 12 seven-branes in an open universe. To close
the universe one needs 24 seven-branes.

   The solution has however the following purely gravitational
interpretation. One considers the metric 
\ben 
ds^2 = g_{ij} \, dy^i
dy^j + e^{\phi}\, dz \, d{\bar z}\,, \label{stringy} 
\een 
where $g_{ij}$
is the previous unimodular metric on the torus $T^2$ with
coordinates $y^i$:
\ben 
g_{ij} = \pmatrix{ \tau_2^{-1}& \tau_1
\tau_2^{-1}\cr \tau_1 \tau_2^{-1}& \tau_1^2 \tau _2 ^{-1} + \tau
_2 \cr }\,. 
\een
This differs from our previous expression (\ref{alg}) essentially by a
conformal transformation, \ie a holomorphic coordinate
transformation on the base metric of this elliptic fibration.

   The metric (\ref{stringy}) is self-dual, or hyper-K\"ahler. If one
takes 24 seven-branes one gets an approximation to a K3 surface
elliptically fibred over $\CP^1$. Essentially this
suggestion is used in \cite{Ooguri-Vafa} to give an interpretation
in terms of D-Instantons. This  proposed construction of certain
limits of  K3 metrics has been  vindicated mathematically  in
\cite{Gross-Wilson}.

\subsection{The ${\overline D}_4$ metric} 

    In this section we shall discuss the ALG metrics in the special case
when $n=4$. The metric is then asymptotic to the flat metric on ${\Bbb
C} \times T^2/{\Bbb Z}_2 $. We shall begin by describing an orbifold
model obtained by setting $m_j=0$, $j=1,2,3,4$, in the asymptotic
metric.  We shall go on to discuss deformations of the orbifold, and
how the orbifold might be blown up. Then we shall discuss an approach
to the metric based on the $su(\infty)$ Toda equation.

\subsubsection{The orbifold model}

   The following description owes
much to a lecture by  Kronheimer several years ago. As far as we are
aware, Kronheimer's work has not appeared explicitly in print.
There is considerable overlap with the papers of Cherkis and
Kapustin.

   We begin by taking ${\Bbb C} \times T ^2$ with complex co-coordinates
$(z, y)$, where $y=y_1+\tau \, y_2$.  The points $(z, \ y)$ and $(z, \
y + R n_1 + i R n _ 2 )$ with $n _ 1, \ n _ 2 \ \in {\Bbb Z}$ and $R
\in {\Bbb R}$ are to be identified.  The metric is obtained by setting
$\tau= {\rm constant} $ in $(\ref{alg})$. Note that to specify the
metric we need three real parameters, namely $R$ and $\tau$, which
characterise the size and shape respectively of the 2-torus.

  We now quotient by the holomorphic involution
\be 
\Gamma: (z, y) \rightarrow (-z, -y)\,, 
\ee
which in polar coordinates defined by $z = r e ^ {i \theta}$ with $0 < r <
\infty$,   $ 0 \le \theta < 2 \pi$, becomes
\be
\Gamma : (r, \theta, y_1 , y_2) \rightarrow (r, \theta + \pi,
-y_1 , -y_2)\,. 
\ee
The involution $\Gamma$ does not
act freely, but rather has 4 fixed points:
\be
(0,0,0,0)\,, \quad  (0,0, { R \over 2}, 0)\,, \quad (0, 0, 0, { R \over
2} ) \,, \quad (0,0, {R \over 2},{R \over 2} )\,. 
\ee
Thus ${\cal M} ^ {\rm sing} ={\Bbb C} \times T ^ 2 / \Gamma$ has 4
singular points, each locally being isomorphic to ${\Bbb C} ^ 2 /
(\pm 1)$. At the fixed points $\Gamma$, being holomorphic, acts as a
self-dual rotation, and so it leaves invariant anti-self-dual
2-forms and negative chirality spinors. Consequently ${\cal M} ^ {\rm
sing}$ is locally flat, but it has holonomy given by $\Gamma$ and
hence may be thought of as having distributional self-dual
curvature localised at the 4 singular points.

   The covering space ${\Bbb C } \times T ^ 2$ is a trivial torus
bundle over $\Bbb C$. The quotient manifold ${\Bbb C} \times T^2
/ \Gamma ={\cal M} ^ {\rm sing}$ may also be thought of as a torus
bundle.  In the language of algebraic geometry ${\cal M} ^ {\rm
sing}$ admits a (singular) elliptic fibration by elliptic curves
\be
\matrix
{T ^ 2 & \longrightarrow & M & {f \atop \longrightarrow} & 
       \cal{M} ^ {\rm sing}\cr
& & \downarrow \pi & & \downarrow \pi _ f\cr
& &{\Bbb C} & & B\cr}\,,
\ee
where the projection map is $\pi: (z, y) \rightarrow (z, 0)$.  The
base space of the induced fibration of $\cal {M}^{\rm sing}$ is ${\Bbb
C} / (\pm1)$, since $(\Gamma, \theta)$ and $(r, \theta + \pi)$ must be
identified; \ie it is a cone with deficit angle $\pi$.  The singular
fibre lying above the vertex of the cone $z = 0$ is not a torus but a
tetrahedron with vertices at $y_1 = 0, y_2=0$; $ y_1 = \ {R \over 2},
y_2=0$; $ y_1 ={R \over 2}, y_2= { R \over 2}$. The metric on the
tetrahedron is flat except at these vertices where there is a deficit
angle of $\pi$.  At infinity the geometry of a large boundary surface
of fixed radius is $(S^1 \times T^2) /(\pm 1)$, \ie we have a twisted
torus bundle over $S^1$.

     The continuous isometries of $\cal {M} ^ {\rm sing}$ consist of
the isometries of ${\Bbb C} \times T ^ 2$ that commute with
$\Gamma$.  This leaves only rotations around the vertex of the cone,
\be
\theta \rightarrow \theta + \ \rm {constant}\,,
\ee
with Killing vector ${\bf m} = {\partial / \partial \theta}$.
By contrast the Killing vectors ${\partial /
\partial y_1 }$ and ${\partial /
\partial y_2}$, which generate translations in the torus,  {\sl do
not}  commute with $\Gamma$.

\subsubsection{The resolved  solution}

   We now consider ``physical picture'' of the blown up metric in the
sense of Page \cite{Page2}, who elaborated a construction first
described in \cite{Gibbons-Pope} by which the K3 metric is built up
from the orbifold $T^4/ (\pm 1)$ with its 16 singular points.  Each
singular point is locally like ${\Bbb C}^2 /(\pm 1)$.  One may ``blow
up'' these singular points, replacing them by copies of ${\Bbb C}
{\Bbb P}^1$.  Now the blow up of ${\Bbb C}^2 / (\pm 1)$ is the
cotangent bundle of ${\Bbb C}{\Bbb P}^1$, \ie $T^{ \star} ({\Bbb
C}{\Bbb P}^1)$, and this carries the self-dual Eguchi-Hanson metric.
To specify the Eguchi-Hanson metric one needs to give three real
parameters, comprising one length scale and two orientations.
Equivalently, one must give a self-dual 2-form in ${\Bbb C} ^2$. This
contributes $16\times 3=48$ parameters in total.  In addition, to
specify the torus one needs to give a further 10 real parameters,
making $48+10=58$ in all.  On the other hand, the specification of a
self-dual metric on K3 requires 58 real parameters, and thus the
counting makes it plausible that the metric on K3 may be approximated
by replacing a small spherical neighbourhood of each singular point by
an Eguchi-Hanson manifold.

   This physical picture has been vindicated by subsequent work by
mathematicians (see \cite{Kobayashi1} for a review). The passage from
the smooth K3 surface to the orbifold limit is referred to as a ``type
I degeneration,'' and convergence is shown in the Gromov-Hausdorff
topology.  

    Let us now apply the idea to the ${\overline D}_4 $ orbifold
${\Bbb C}^2 \times T^2 / (\pm 1)$, which may be regarded as a limit of
the $T^4 / (\pm 1)$ orbifold.  There are 3 real parameters for $T^2$
and $3 \times 4 = 12$ for the four Eguchi-Hanson metrics, making 15 in
all.  Each ${\Bbb C}{\Bbb P}^1$ has self-intersection number $+2$, and
the torus gives a fifth homology 2-cycle, which intersects each of the
four $ {\Bbb C}{ \Bbb P}^1$'s at one point. The intersection form is
therefore given by the extended Dynkin diagram $\overline {D}_4$: the
rank and signature are both equal to five, and therefore $b_2^+ =
5$. This agrees with our parameter count, since the number of zero
modes of the Lichnerowicz operator is $3 \ b_2^+=15$ \cite{hawpop}.
Since the metric is self-dual there are no negative modes, which
suggests that these configuration are at worst neutrally stable.

\section{The Toda equation} 

    The metrics (\ref{alg}) are flat at infinity, and since 
the general ideas of Kaluza-Klein theory suggest that
deviations from flatness are governed by the Laplace operator with
a mass term, the approach to flatness should be exponentially fast.
As mentioned above,  the metric cannot be expected to have more
than one Killing vector. In the orbifold case, ${\bf m} = 
m^\alpha {\partial /
\partial x^\alpha} = {\partial / \partial \theta}$ is a Killing vector,
and this can be expected to survive.  In terms of the coordinates
$z$ and $y$,  $\partial /\partial \theta$ is holomorphic, but
this complex structure, let us call it $\eta^3$, is privileged.  The
Killing vector $\partial / \partial \theta$ is {\bf {not}}
tri-holomorphic; the $U(1)$ it generates will rotate the 2
orthogonal complex structures $\eta^1$ and $\eta^2$ into
each other.   Thus the Killing vector $m_\alpha$ is anti-self-dual: 
$m_{ [\alpha ; \beta]} = - \star m _{[\alpha; \beta]}$.  It
follows that the metric may be cast in the form \cite{boyfin,dasgeg}
\be 
ds^2 = {\dot \nu} ^ {-1}\, (2d \theta + \nu_u\, dv - \nu_v\,  d u)^2 
+ \dot {\nu} \, [dt^2 + e^\nu\, (du^2 + dv^2)]
\label{todamet} \,,
\ee 
where $\nu$ satisfies the Toda equation 
\be
(e^\nu)^{\cdot \cdot} + \nu_{uu} + \nu_{vv} = 0\,,
\label{todaeqn} 
\ee 
with $\cdot$ denoting ${\partial /
\partial t}$.  The interpretation of the coordinate $t$ is
that it is the moment map associated to ${\partial /
\partial \theta}$ regarded as a Hamiltonian vector field with respect to the
privileged symplectic structure $\omega^3$.  The complex
coordinate $w = u + \im\, v $ parameterises the symplectic quotient of
the 4-manifold by the $U(1)$ action, thus: 
\be 
\omega^3 = {\nu}\,  e^\nu \, du \wedge dv - 
   dt \wedge (2d \theta + \nu_u \, d v - \nu_v\,  du)\,, 
\ee
and  
\be 
{\cal L}_m \eta ^3 = 0 \Longrightarrow  i_{\partial \over \partial
\theta} \, \eta ^3 =2dt\,.
\ee
Note that the closure of $\eta ^3$ is
equivalent to the Toda equation (\ref{todaeqn}). 

   The geometric picture is as follows. Locally, the manifold is
foliated by level sets $t={\rm constant}$. The orbits of the $SO(2)$
action lie in these level sets. The coordinates $(u,v)$ or $(y_1,
y_2)$ parameterise the two-dimensional space $\Sigma_2 \equiv T^2$ of
orbits. The symplectic form $\eta^3 $ descends to give a symplectic
form, or area form, on the symplectic quotient $\Sigma_2$.  The
freedom to choose canonical or Darboux coordinates on the quotient
gives rise to the gauge group $SDiff(T^2)$, whose Lie algebra
$sdiff(T^2)$ is also known as $su(\infty)$, $A_\infty$ or $w_\infty$.

   To summarise the above discussion, we saying that the exact
non-singular ALG metrics, whose asymptotic forms are given by
(\ref{alg}), will, in certain special cases, admit a
non-triholomorphic circle action, and that such ALG metrics  
must necessarily be contained within the class of metrics
(\ref{todamet}), where $\nu$ satisfies the Toda equation
(\ref{todaeqn}).  Although explicitly solving the Toda equation is
difficult we may, nevertheless, be able to use it in order to make a
large-distance perturbative analysis.  Thus we may take an
explicit large-distance solution of the form (\ref{alg}) as a
zeroth-order starting point, re-express it in the Toda metric form
(\ref{todamet}), and then look for perturbations around it that
satisfy the Toda equation (\ref{todaeqn}).   The asymptotic solution
of the form (\ref{alg}) that provides the zeroth-order starting point
must be one that is expected to extend to an exact solution with a
non-triholomorphic circle action.  Two natural candidates present
themselves, namely the solutions with the holomorphic function $\tau(z)$ 
given by (\ref{potxxx}) for $n=0$, or for $n=4$ with $\overline
m_j=0$.  This latter example is in fact nothing but the flat metric,
and we shall study it first.

\subsection{Perturbation around the $n=4$ flat metric}

   This first example is obtained by setting $n=4$ in (\ref{potxxx}),
and taking the parameters $\overline m_j$ all to be zero.  This
implies that $\tau(z)$ will be a constant.  By making a
convenient choice of moduli for the resulting metric, in which we set $\tau=
\im$ to get a square torus, we see that the metric (\ref{alg}) becomes
\be
ds^2 = dy_1^2 + dy_2^2 + dz\, d\overline z\,.\label{flatxxx}
\ee

  Within the Toda class of metrics (\ref{todamet}), the flat metric
corresponds to taking
\be
e^\nu = A \, t\,,
\ee
where $A$ is a constant which may, by suitable rescaling of $u$ and
$v$, be set to $4$. We may then identify $2(u,v)$ in (\ref{todamet})
with $(y_1,y_2)$ in (\ref{flatxxx}). The metric (\ref{todamet})
becomes
\be 
ds^2 = 4t\, 
\ d \theta^2 + {1 \over t}\,  dt^2 + |dy|^2 \,.
\ee 
The radial
proper distance is  $\rho = 2 \sqrt {t}$, and if $0 \leq \theta < 2
\pi$, we get a flat metric on a cone with deficit angle $\pi$.
Writing 
\be 
e^\nu = 4t + \lambda\,, 
\ee
where $\lambda$ is small
compared with $t$, we find that (\ref{todaeqn})   becomes 
\be 
t \
\lambda_{tt} + \lambda_{uu} + \lambda_{vv} = 0\,. 
\ee
If we make the ansatz
\be 
\lambda = f(t)\,  \exp \im\,  (k_1\,  u + k_2\, v)\,,
\ee
then if $\lambda$ is small
compared with $t$ near infinity, the regular 
solution takes the form $f\sim\, \rho\, K_1(k\, \rho)$ where $K_1$ is
a modified Bessel function, and so asymptotically we have 
\be 
f \sim (\hbox{const})\,  \rho^{1\over 2} \,  e^{- \rho\, r}\,, 
\ee 
where
\be k\equiv  \sqrt{k_1^2 + k_2^2 } \,.
\ee
Thus indeed the deviations from flatness, which necessarily entail
a $U(1) \times U(1)$ symmetry violation, fall off exponentially as claimed
above.

  The terms $(\nu_u \, dv - \nu_v \, du)$ in the metric (\ref{todamet}) 
correspond to the presence of magnetic fields along the $x^3$
direction.  These fall off exponentially away from the core
region.

   Later, in section \ref{n0case}, we shall discuss the $n=0$ metric 
in the class described by (\ref{alg}) with $\tau$ given by
(\ref{tauchoice}).

\subsection{Metrics with both self-dual and 
anti-self-dual Killing vectors}\label{sdasd}

  In this case the metric depends upon a free function of two
variables.  It is easy to specify this function if one uses the
harmonic function description adapted to the tri-holomorphic Killing
vector $ \partial / \partial \psi$ say.  It appears to be difficult to
translate this explicitly into the solution of the Boyer-Finley
\cite{boyfin,dasgeg} equation depending upon two variables that one
obtains if one uses the Toda description adapted to the
mono-holomorphic Killing vector $\partial / \partial \theta$ and
imposes translational symmetry.

   Suppose the metric with the tri-holomorphic Killing vector is also
axisymmetric, and thus it takes the form 
\be 
ds^2 = V^{-1} \,(d \psi +
\omega \, d \theta )^2 + V ( d z^2 + d \rho^2 + \rho^2\,  d \theta^2
)\,, \label{sdasdmet}
\ee
where $V$ and $\omega$ depend only on $\rho$ and $z$. We have 
\be
\rho V_z = \omega_\rho, \qquad \rho \, V_\rho =-\omega_z\,.
\label{CR1} 
\ee
The privileged symplectic form is 
\be 
\eta^3= (d \psi + \omega \, d
\theta ) \wedge  dz + V \, \rho \, d \theta  \wedge d \rho\,,
\label{expression}
\ee
and is
closed by virtue of the second equation in (\ref{CR1}). The
closure guarantees the local existence  of the moment map $t$
which satisfies  
\be 
i_{\partial \over
\partial \theta} \, \eta^3 = 2dt = \omega \, dz + \rho \, V \, d \rho\,.
\ee
Thus 
\be 
2t_z=\omega\,, \qquad  2t_\rho = \rho\,  V.
\ee
whence
Note that adding a constant $c$ to $V$ shifts $t$ by $c\, \rho$. 
We also have another closed one-form given by
\be
2du= V dz - {\omega \over \rho} d \rho.
\ee

If we now set $e^\nu =\rho ^2$ and use the chain rule we find that
$\nu =\nu( t, u)$ will satisfy the Toda equation with no $v$
dependence.
If we set $2v=\tau$ we can pass between the two forms of the metric.
In particular
\be
{\partial \nu \over \partial t}= {2 V \over \omega ^2 + \rho ^2 V^2 } 
\ee
and
\be
{\partial \nu \over \partial u}= {2 \omega \over \omega ^2 + \rho ^2 V^2}. 
\ee

   A different approach \cite{Ward} to relating solutions of
the Toda equation to harmonic functions starts with an axisymmetric
harmonic function $H(\rho, z)$. 
If 
\be
T= \ft12 \rho\,  H_\rho\,, \qquad x=-H_z,
\ee
and $f=\log(\rho^2/4)$ then we get a translation-invariant
solution of the Toda equation 
\be (e^f)_{TT} + f_{xx}=0\,.  
\ee

   One may reverse the steps. Given $f$, one may obtain $H$ and then
construct a metric with triholomorphic $U(1)$ by setting $V=1+H_z$ and
$\omega=-\rho H_\rho$.  Note that starting with a given $V$ and
$\omega$ one gets in general two different solutions of the Toda
equation. Only the first approach leads to the conventional Toda form
of the metric and the moment map $t$.\footnote{We thank Paul Tod and
Richard Ward for helpful discussions on this and related points.}

\subsubsection{The $n=0$ Cherkis-Kapustin metric}\label{n0case}

    After appropriate rescalings and elimination of inessential
constants, in this case we can take the holomorphic function $\tau(z)$ in
(\ref{alg}) to be given by (\ref{tauchoice}) with $n=0$, implying
\be
\tau_1 = \theta\,,\qquad \tau_2=\log r\,.
\ee
After the transformation (\ref{legendre}), and converting to the
notation of (\ref{sdasdmet}) the metric (\ref{alg}) then becomes
\be
ds^2 = \fft{1}{\log \rho}\, (d\psi -z\, d\theta)^2 + \log \rho\, (dz^2 +
d\rho^2 +\rho^2\, d\theta^2)\,,
\ee
with $V=\log\rho$ and $\omega=-z$.

   The transformation to coordinates of the Toda metrics
(\ref{todamet}) can be effected by noting from (\ref{CR1}) that we
have the two exact 1-forms
\bea
dt &=& w\, dz+ V\, \rho\, d\rho =-z\, dz + \rho\, \log\rho\,
d\rho\,,\nn\\
du &=& V\, dz - \fft{\omega}{\rho}\, d\rho= \log\rho\, dz -
\fft{z}{\rho}\, d\rho\,,
\eea
which can be integrated to give 
\be
t= -\ft12 z^2 +\ft14 \rho^2\, (2\log\rho -1)\,,\qquad
u= z\, \log\rho\,.
\ee
We also make the identification $2v=\tau$.
It is easily verified that the function $\nu$ in the Toda form of the
metric is given by
\be
e^\nu = \rho^2\,,
\ee
and, by use of the chain rule, that this indeed satisfies the Toda
equation (\ref{todaeqn}).

 \subsubsection{The Calderbank-Tod-Nutku-Sheftel solution}

  This metric \cite{Calderbank-Tod,Nutku-Sheftel} is a solution of the
Toda or Boyer-Finley equation depending upon two holomorphic functions
$a(u+\im\, v)$ and $b(u+\im\, v)$, one of which is essentially the
gauge freedom to make holomorphic coordinate transformations of
$u+\im\, v$. Thus in effect it depends on one analytic function, or
one free function of two variables, which is just the freedom in
giving one of the metrics in the previous section.  However the two
solutions are distinct \cite{Calderbank-Tod}.  The
Calderbank-Tod-Nutku-Sheftel solution is
\be 
e^\nu ={|(t+ a(u+\im\, v))\, b^{\prime}  (u+\im\, v) |^2 \over (1+
|b(u+\im\, v) |^2 )^2 }\,. 
\ee

\subsubsection{Separable solutions}

   These are of the form
\be
e^\nu= (t^2 + 2\beta\, t +\gamma)\,  e^\phi(u,v)\,,
\ee
where $\beta$ and $\gamma$ are constants, and $\phi$ is a solution of 
the Liouville equation
\be
(\partial_u \partial_u +\partial_v \partial_v)\,  \phi =2 e^\phi\,. 
\ee
The general solution of Liouville's equation leads to
\be
e^\nu={ (t^2 +2\beta\, t +\gamma)\, |b^\prime(u+iv)|^2 
      \over (1+|b(u+iv)|^2 ) ^2 }, 
\ee
where $b(u+\im\, v)$ is again an arbitrary holomorphic function.
As before, the freedom to change coordinates means that
one might as well set $b(u+\im\, v)=u+iv$, and shifting $t$ allows one to set 
$\beta=0$.  One gets the Eguchi-Hanson metric in this way.

\subsubsection{Further solutions}

   Other known solutions have been recast in the Toda form, including
the Atiyah-Hitchin metric.  Because the $SO(3)$ isometry group does not act
tri-holomorphically, there is a solution of the Toda equation for
every $SO(2)$ subgroup. These have been worked out by Bakas and
Sfetsos \cite{Bakas-Sfetsos} and may be useful in suggesting an
ansatz for the more complicated ALG case.

  As explained in \cite{Cherkis} and noted earlier by Kronheimer and
Nakajima, one may obtain ALG metrics from the modulus spaces of
solutions of the Hitchin equations on certain Higgs bundles over
surfaces. The latter may be considered as the dimensional
reduction of the self-dual Yang-Mills equations on ${\Bbb E}^2
\times S$, where $S$ is a Riemann surface.

\subsection{Some properties of the Toda system: Lax pair
and spinors}

   We recall a few useful properties of the Toda
equation. Firstly, we may introduce a quantity $\phi$ by $\nu=
{-\ddot \phi}$. The equation (\ref{todaeqn}) then becomes 
\be
\partial {\overline \partial } \phi= \exp (- {\ddot \phi})\,.
\ee
This is manifestly of Toda form if one thinks of the operator
$-{\partial  ^2 \over \partial t^2 }$ as the limit to zero step-size 
of the leap-frog difference operator $-\phi_{n-1}  +2
\phi_{n} - \phi_{n+1} $. Considered as an infinite matrix, this is
the Cartan matrix $A_{\infty}$ which is associated to the Lie
algebra $su(\infty)$.

   Secondly, a Lax pair may be constructed by considering
two time-dependent operators $L$ and $N$ acting on ${\Bbb
C}$-valued  functions $f= f(y_1, y_2)$ :
\bea 
N f &=& (\partial \nu)f \,,\nn\\
L f &=& {\overline \partial } f + ({\dot \nu }\, e^\nu ) \, f\,. 
\eea
One has
\be 
[N,L]f= -(\partial {\overline \partial } \nu ) f\,, 
\ee
and hence 
\be
 {\dot L} = [N,L]\,. 
\ee
The operator $L$ is a modified Dirac operator acting on spinors
defined over the 2-dimensional quotient manifold $\Sigma_2 \equiv
T^2$. 


\section{Conclusions}

    In this paper, we have studied several examples of new metrics
with special holonomy, and their significance in string theory and
M-theory.  We first considered the $\bC_8$ metrics whose existence was
demonstrated by numerical analysis in \cite{g2spin7}.  They have
cohomogeneity one, with $S^7$ principal orbits, described as
triaxially-squashed $S^3$ bundles over $S^4$, which degenerate to
$\CP^3$ on a bolt at short distance.  The $\CP^3$ itself is viewed as
an $S^2$ bundle over $S^4$, and a non-trivial parameter $\lambda$ in
the $\bC_8$ metrics characterises the squashing of the $\CP^3$.
Regular metrics arise for $\lambda^2\le4$, with the limiting case
$\lambda^2=4$ being the standard asymptotically conical (AC) Ricci-flat
K\"ahler Calabi metric
on the complex line bundle over the Fubini-Study metric on $\CP^3$.
For $\lambda^2<4$ the metrics have Spin(7) holonomy, and they are
asymptotically locally conical (ALC). At large distance they are locally of
the form of a product of a circle of stabilised radius and an AC
7-metric of $G_2$ holonomy. The metric functions in the three
directions in the $S^3$ fibres in $\bC_8$ behave similarly to those in the
Atiyah-Hitchin metric, and we find that this leads to a natural
orientifold picture for an M-theory solution (Minkowski)$_4\times
\bC_8$ reduced on the asymptotic circle, with D6-branes wrapped over
$S^4$.

    We next considered a new class of complete $G_2$ metrics, which
are of cohomogeneity one with $S^3\times S^3$ principal orbits.  By
starting from a short-distance Taylor expansion, and then studying the
evolution to large distances numerically, we established that there
exist regular metrics, which we denote by $\bC_7$, in which the
principal orbits degenerate to a $T^{1,1}$ bolt at short distance.
The metrics are ALC, locally approaching the product of a stabilised circle 
and the deformed six-dimensional conifold over $T^{1,1}/Z_2$ at large
distance.  An intriguing feature of these metrics is that the circle
whose radius stabilises at large distance also remains non-singular at
short distance.  Thus if we perform a Kaluza-Klein dimensional
reduction of an M-theory solution (Minkowski)$_4\times \bC_7$ on this
circle, we get a type IIA solution with an everywhere non-singular
dilaton.   

   We then considered a new class of four-dimensional hyper-K\"ahler 
metrics described by Cherkis and Kapustin \cite{Cherkis}, following
earlier related work by Kronheimer.  Only the asymptotic form of these
metrics is known explicitly; at large distance they are
characterised by a holomorphic or anti-holomorphic function, and they
admit a tri-holomorphic $T^2$ action.  In the interior, the metrics
will not in general admit any Killing symmetries at all, although in
special cases there can be a mono-holomorphic $U(1)$ symmetry.  In
such cases the metric falls into a class described in terms of a
function satisfying the $su(\infty)$ Toda equation, and this can
provide a way of studying the behaviour of the metric in the interior
region.

\section*{Acknowledgements}

    We are grateful to Nigel Hitchin, Peter Kronheimer, Paul Tod and
Richard Ward for discussions, and to James Sparks for pointing out an
error in our previous interpretation of the type IIA solution in
section 3.2.3.  We thank Anton Kapustin for a helpful correspondence
drawing our attention to the possibility that the exact 2-monopole
modulus space might not admit an exact circle action, and for drawing
our attention to the revision of \cite{Cherkis}.  M.C. is supported in
part by DOE grant DE-FG02-95ER40893 and NATO grant 976951; H.L.~is
supported in full by DOE grant DE-FG02-95ER40899; C.N.P.~is supported
in part by DOE DE-FG03-95ER40917.  M.C. is grateful for hospitality in
the Physics Department at Rutgers University during the course of this
work.

\bigskip\bigskip
\appendix
\centerline{\Large\bf Appendices}

\section{Atiyah-Hitchin metric}\label{ahback}

   In this appendix, we review some basic results about the
Atiyah-Hitchin metric.  We begin in section \ref{ahsolution} with a
review of the solution itself, as discussed in \cite{atihit}.  Then,
in section \ref{ahglobal}, we give a discussion, based on one in
\cite{gibman}, of the global structure of the Atiyah-Hitchin metric,
focusing especially on the identifications of the angular coordinates on the
principal orbits.

\subsection{The solution}\label{ahsolution}

   We write the metric in the form
\be
ds_4^2 = dt^2 + \ft14(a_1^2\, \sigma_1^2 + a_2^2\, \sigma_2^2
+ a_3^2\, \sigma_3^2)\,,\label{metric}
\ee
where the metric functions satisfy the first-order equations
\be
\dot a_1 = \fft{a_1^2 - (a_2-a_3)^2}{a_2\, a_3}\,,\quad
\dot a_2 = \fft{a_2^2 - (a_3-a_1)^2}{a_3\, a_1} \,,\quad
\dot a_3 = \fft{a_3^2 - (a_1-a_2)^2}{a_1\, a_2}\,.\label{ahfo}
\ee
These can be solved by defining a new coordinate
$\tau$, related to $t$ by $dt= -\ft18(a_1\, a_2\, a_3)\,  u^{-2}\,
d\tau$, with $u$ being a solution of
\be
\fft{d^2u}{d\tau^2} + \ft14 u\, \cosec^2 \tau  = 0\,.\label{elleq}
\ee
It can then be verified that the solution is given by
\be
a_1 = 2\sqrt{\fft{w_2\, w_3}{w_1}}\,,\qquad
a_2 = 2\sqrt{\fft{w_3\, w_1}{w_2}}\,,\qquad
a_3 = -2\sqrt{\fft{w_1\, w_2}{w_3}}\,,
\ee
where
\be
w_1 = -u\, u' - \ft12 u^2\, \cosec \tau\,,\quad
w_2 = -u\, u' + \ft12 u^2\, \cot \tau\,,\quad
w_3 = -u\, u' + \ft12 u^2\, \cosec \tau\,,
\ee
and $u'$ means $du/d\tau$.  The solution of (\ref{elleq}) is taken to be
\be
u= \fft1{\pi}\, \sqrt{\sin \tau}\, K(\sin^2\ft12 \tau)\,,
\ee
where
\be
K(k) \equiv \int_0^{\pi/2} \fft{d\phi}{(1-k\, \sin^2\phi)^{1/2}}
\ee
is the complete elliptic integral of the first kind.  The coordinate
$\tau$ ranges from $\tau=0$ at the bolt, to $\tau=\pi$ at infinity.

    Near $\tau=0$ we have
\bea
a_1 &=& \ft18 \tau^2 +\ft1{192} \tau^2+\cdots\,,\nn\\
a_2 &=&  1+\ft1{32} \tau^2+\cdots\,,\nn\\
a_3 &=& -1+ \ft1{32} \tau^2+\cdots\,.
\eea
In terms of the proper-distance coordinate $t=\ft1{32} \tau^2+\cdots$,
we see that near the bolt we shall have
\be
ds_4^2 \sim dt^2 + 4t^2\, \sigma_1^2 +
\ft14 (\sigma_2^2 + \sigma_3^2)\,.\label{ahshort}
\ee

   Near $\tau=\pi$, which is the asymptotic region at infinity, one finds
that $a_1$ and $a_2$ grow linearly with proper distance, while $a_3$ tends
to a constant.  To see this, we use the facts that
\bea
K(1-\ep) &\sim& -\ft12 \log\ep -\log 2 + O(\ep\, \log\ep)\,,\nn\\
E(1-\ep) &\sim& 1-\ft14 (\log\ep + 2\log 2)\, \ep + O(\ep^2\, \log\ep)
\,,
\eea
as $\ep$ goes to zero, where $E(k)=\int_0^{\pi/2}\, \sqrt{1-k\,
\sin^2\phi}\, d\phi$ is the complete elliptic integral of the second
kind.  In particular, we see that $a_3$ approaches the value
\be
a_3(\infty) = -\fft{2}{\pi}
\ee
at infinity, while $a_1$ and $a_2$ have the asymptotic form $a_1\sim
a_2\sim \pi^{-1}\, \log\ep$, where $\ep=\cos^2\ft12 \tau$.  We also
find $u\sim -(\sqrt2\, \pi)^{-1}\, \ep^{1/4}\, \log\ep$, and hence
$\ep\sim e^{2\pi\, t}$.  At leading order, the metric
at infinity therefore approaches
\be
ds_4^2 \sim dt^2 + t^2\, (\sigma_1^2 + \sigma_2^2) + \fft1{\pi^2}\,
\sigma_3^2\,.
\ee

\subsection{Global considerations}\label{ahglobal}

   We shall represent the left-invariant 1-forms $\sigma_i$ in terms of
Euler angles
\bea
\sigma_1 &=& -\sin\psi \, d\theta + \cos\psi\, \sin\theta\,
d\phi\,,\nn\\
\sigma_2 &=& \cos\psi \, d\theta +\sin\psi\, \sin\theta\,
d\phi\,,\nn\\
\sigma_3 &=& d\psi + \cos\theta\, d\phi\,.\label{euler}
\eea
These are the natural ones to use in the asymptotic region
near infinity, where it is $\sigma_3^2$ that has a coefficient tending
to a constant.\footnote{The roles of $\sigma_1$ and $\sigma_2$ are
reversed compared with our usual conventions.  This is in order to
maintain the same conventions as are customarily used when discussing
the Atiyah-Hitchin metric.   In particular, we now have
$d\sigma_i=+\ft12 \ep_{ijk}\, \sigma_j\wedge\sigma_k$.}
However, near the bolt, where the coefficient of $\sigma_1^2$ goes to
zero, it is more natural to consider a redefined set of 1-forms
$\td\sigma_i$, where
\be
\sigma_1 =\td\sigma_3\,,\qquad \sigma_2=\td\sigma_1\,,\qquad
\sigma_3=\td\sigma_2\,.\label{sigtdsig}
\ee
These can be represented in terms of tilded Euler angles:
\bea
\td\sigma_1 &=& -\sin\wtd\psi \, d\td\theta + \cos\wtd\psi\, \sin\td\theta\,
d\td\phi\,,\nn\\
\td\sigma_2 &=& \cos\wtd\psi \, d\td\theta + \sin\wtd\psi\, \sin\td\theta\,
d\td\phi\,,\nn\\
\td\sigma_3 &=& d\wtd\psi + \cos\td\theta\, d\td\phi\,,\label{tdeuler}
\eea
An explicit relationship between the tilded and untilded Euler angles
can be given, but we do not need that here.

   It is clear from (\ref{ahshort}) that regularity near the bolt
requires that $\wtd\psi$ should have period $\pi$.  We now need to
express this in terms of the original untilded Euler angles, which we
shall be using in the asymptotic region.  To do this, let us begin by
assuming that the $\sigma_i$, and hence also the $\td\sigma_i$, are
left-invariant 1-forms on $SO(3)$ rather than $SU(2)$, meaning that
the periods of the fibre coordinates will be $\Delta\psi=2\pi$, and
also $\Delta\wtd\psi=2\pi$.  We see that this means we should identify
$\wtd\psi$ further, under a discrete transformation $\wtd I_3$ defined by
\be
\wtd I_3:\qquad \td\theta\longrightarrow \td\theta\,,\quad
           \td\phi \longrightarrow \td\phi\,,\quad
           \wtd\psi \longrightarrow \pi+\wtd\psi\,.
\ee
Observe that $\wtd I_3$ has the following action on the three $\td\sigma_i$
1-forms:
\be
\wtd I_3(\td\sigma_1)=-\td\sigma_1\,,\qquad
\wtd I_3(\td\sigma_2)=-\td\sigma_2\,,\qquad
\wtd I_3(\td\sigma_3)=+\td\sigma_3\,.
\ee
Note that this leaves the algebra, $d\td\sigma_i = \ft12\ep_{ijk}\,
\td\sigma_j\wedge \td\sigma_k$, invariant.
One can also define permuted analogues of this discrete operation,
which instead leave either $\td\sigma_1$ or $\td\sigma_2$ fixed in
sign, while reversing the signs of the other two.  These are defined
by
\bea
&&\wtd I_1(\td\sigma_1)=+\td\sigma_1\,,\qquad
\wtd I_1(\td\sigma_2)=-\td\sigma_2\,,\qquad
\wtd I_1(\td\sigma_3)=-\td\sigma_3\,,\nn\\
&&\wtd I_2(\td\sigma_1)=-\td\sigma_1\,,\qquad
\wtd I_2(\td\sigma_2)=+\td\sigma_2\,,\qquad
\wtd I_2(\td\sigma_3)=-\td\sigma_3\,.
\eea
It can be seen from (\ref{tdeuler}) that the corresponding actions of
$\wtd I_1$ and $\wtd I_2$ on the tilded Euler angles will be
\bea
\wtd I_1:&& \td\theta\longrightarrow \pi-\td\theta\,,\quad
            \td\phi\longrightarrow \pi+\td\phi\,,\quad
            \wtd\psi\longrightarrow -\wtd\psi\,,\nn\\
\wtd I_2:&& \td\theta\longrightarrow \pi-\td\theta\,,\quad
            \td\phi\longrightarrow \pi+\td\phi\,,\quad
            \wtd\psi\longrightarrow \pi-\wtd\psi\,.
\eea

   We want to see how $\wtd I_3$ acts on the untilded Euler angles.
Since the tilded and untilded quantities are related by permutation,
as in (\ref{sigtdsig}), it follows that there will be an identical
permutation relation for the discrete operators, namely
\be
I_1 =\wtd I_3\,,\qquad I_2=\wtd I_1\,,\qquad
I_3=\wtd I_2\,,\label{itdi}
\ee
where the $I_i$ act on untilded quantities exactly as $\wtd I_i$ act
on tilded quantities.  Thus we have
\bea
I_1:&& \theta\longrightarrow \pi-\theta\,,\quad
            \phi\longrightarrow \pi+\phi\,,\quad
            \psi\longrightarrow -\psi\,,\nn\\
I_2:&& \theta\longrightarrow \pi-\theta\,,\quad
            \phi\longrightarrow \pi+\phi\,,\quad
            \psi\longrightarrow \pi-\psi\,,\nn\\
I_3:&& \theta\longrightarrow \theta\,,\quad
           \phi \longrightarrow \phi\,,\quad
           \psi \longrightarrow \pi+\psi\,.
\eea

    In particular, we see that the identification $\wtd I_3$ that we
needed in order to get a metric regular near the bolt implies, in
terms of the untilded coordinates that it is natural to use near infinity,
that we must identify under $I_1$, namely
\be
I_1:\qquad \theta\longrightarrow \pi-\theta\,,\quad
            \phi\longrightarrow \pi+\phi\,,\quad
            \psi\longrightarrow -\psi\,.\label{i1id}
\ee

   It can be noted that one is also free to impose in addition an
identification under $I_3$, since this has no fixed points on the
bolt.  But it should be emphasised that the identification under $I_1$
is obligatory, whilst the further identification under $I_3$ is optional.

\subsection{Asymptotic behaviour of the Atiyah-Hitchin metric}\label{ahasymp}

   The first-order equations for the Atiyah-Hitchin system are given
by (\ref{ahfo}).  In terms of a new radial coordinate $r$, defined by
$dt=h\, dr$, a simple known
solution is the Taub-NUT metric, given by
\be
a_1=a_2= 2(r^2-m^2)^{1/2}\,,\quad a_3= 4m\,
\Big(\fft{r-m}{r+m}\Big)^{1/2}\,,\quad h =
\Big(\fft{r+m}{r-m}\Big)^{1/2}\,.
\ee

   We now look for a solution perturbed around the Taub-NUT metric,
in which $a_1$ and $a_2$ are
unequal, but where they approach equality asymptotically at large
$r$.  We make a perturbative expansion
\be
a_1= A_0 + A_1\,,\quad a_2=A_0 + A_2\,,\quad a_3 = B_0 + A_3\,,\quad
h= h_0 + h_1\,,
\ee
where $A_0$, $B_0$ and $h_0$ will now be the zeroth-order solution
above, {\it i.e.}
\be
A_0= 2(r^2-m^2)^{1/2}\,,\quad B_0= 4m\,
\Big(\fft{r-m}{r+m}\Big)^{1/2}\,,\quad h_0 =
\Big(\fft{r+m}{r-m}\Big)^{1/2}\,,\label{zerothah}
\ee
and we shall work to linear order in
the functions $A_1$, $A_2$, $A_3$ and $h_1$.

    We can choose a gauge for $h_1$ such that $A_2=-A_1$, by imposing
\be
h_1 = \fft{h_0\, A_3}{2 A_0 -B_0}\,.
\ee
The linearised equations for $A_1$ and $A_3$ then become
\bea
A_1' &=& \fft{4 h_0\, A_1}{B_0} - \fft{h_0\, B_0\, A_1}{A_0^2}\,,\nn\\
A_3' &=& \fft{h_0\, B_0\, (4A_0 -B_0)\, A_3}{A_0^2\, (2A_0-B_0)}\,.
\eea
A simple solution can be obtained by taking $A_3=0$.  We see that the solution
for $A_1$ is given by
\be
A_1\sim e^{\int( \ft{4h_0}{B_0} - \ft{h_0\, B_0}{A_0^2})}\,.
\ee
From the zeroth-order solution (\ref{zerothah}) we therefore find
that the dominant large-$r$ behaviour for $A_1$ will be
\be
A_1 \sim  e^{\int \ft{4h_0}{B_0}} \sim e^{r/m}\,.
\ee
It follows that our assumption that $A_1$ is a small perturbation will
be valid only if
\be
m<0\,,
\ee
and hence the Atiyah-Hitchin metric is asymptotic to Taub-NUT with
a {\it negative} mass.

   To see this explicitly, consider the M-theory solution
\be
d\hat s_{11}^2 = dx^\mu\, dx_\mu + ds_4^2\,,
\ee
where $dx^\mu\, dx_\mu$ is the metric on 8-dimensional Minkowski
spacetime.  We reduce this to the Einstein-frame metric in $D=10$, using the
standard Kaluza-Klein reduction
\be
d\hat s_{11}^2 = e^{-\fft16\phi}\, ds_{10}^2 + e^{\fft43\phi}\,
(dz+\cA_\1)^2\,.
\ee
The reduction is performed on the circle parameterised by $\psi$ in
(\ref{euler}), we can be done at large distance because $a_1$ and
$a_2$ become equal asymptotically, and so $\del/\del\psi$
asymptotically becomes a Killing vector.  Thus we get
\be
ds_{10}^2 = a_3^{1/4}\, dx^\mu\, dx_\mu + a_3^{1/4}\, (dt^2 +
     a_1^2\, \sigma_1^2 + a_2^2\, \sigma_2^2)\,.
\ee
The metric coefficient $g_{00}$ for $dx^0\, dx^0$ is therefore given
by
\be
g_{00} = -a_3^{1/4}\,.\label{g00}
\ee

  For the Taub-NUT metric we therefore have
\be
g_{00} = (4m)^{1/4}\, \Big( 1 - \fft{m}{4r} + \cdots\Big)\,,\label{tnmass}
\ee
allowing us to read off the mass as $M=\ft14 m$.  Thus our
perturbative discussion above has shown that the mass $M$ is negative
in the Atiyah-Hitchin solution.


\begin{thebibliography}{99}

\bm{hitch} N.J. Hitchin, {\sl Stable forms and special metrics,}
math.DG/0107101.  

\bibitem{g2spin7}
M.~Cveti\v c, G.~W.~Gibbons, H.~L\"u and C.~N.~Pope,
{\sl Cohomogeneity one manifolds of Spin(7) and G(2) holonomy,}
hep-th/0108245.

\bm{atihit}  M.~F.~Atiyah and N.~J.~Hitchin, {\sl Low-energy scattering of
nonabelian monopoles}, Phys. Lett. {\bf A107}, 21 (1985).

\bm{gibman} G.W.~Gibbons and N.S.~Manton, {\sl Classical and quantum 
dynamics of BPS monopoles}, Nucl. Phys. {\bf B274}, 183 (1986). 

\bm{Sen} A.~Sen, {\sl A note on enhanced gauge symmetries 
in M- and string theory,} JHEP {\bf 9709}, 001 (1997), hep-th/9707123.

\bm{brgogugu} A. Brandhuber, J. Gomis, S.S. Gubser and S. Gukov, {\sl
Gauge theory at large $N$ and new $G_2$ holonomy metrics}, hep-th/0106034.

\bm{Cherkis} S.A. Cherkis and A. Kapustin, {\sl New hyper-K\"ahler metrics
from periodic monopoles}, hep-th/0109141.

\bm{boyfin} C.P. Boyer and J.D. Finley III, {\sl Killing vectors in 
self-dual Euclidean Einstein spaces}, J. Math. Phys. {\bf 23}, 1123 
(1982).

\bm{dasgeg} A. Das and J.D. Gegenberg, {\sl Stationary Riemannian space-times
with self-dual curvature}, Gen. Rel. and Gravitation {\bf 16},  
817 (1984).

\bm{cglpnew} M.~Cveti\v c, G.~W.~Gibbons,
H.~L\"u and C.~N.~Pope, {\sl New complete non-compact Spin(7)
manifolds,} hep-th/0103155; {\sl New cohomogeneity one metrics with
Spin(7) holonomy,} math.DG/0105119.

\bm{brysal} R.L. Bryant and S. Salamon, {\sl On the construction of
some complete metrics with exceptional holonomy}, Duke Math. J. {\bf
58}, 829 (1989).

\bm{gibpagpop} G.W. Gibbons, D.N. Page and C.N. Pope, {\sl Einstein
metrics on $S^3$, $\R^3$ and $\R^4$ bundles}, Commun. Math. Phys.
{\bf 127}, 529 (1990).

\bibitem{Calabi} E. Calabi, {\sl Metriques Kahleriennes et fibr\'es 
holomorphes}, Ann. Sci. de l'E.N.S. {\bf 12}, 266 (1979).

\bibitem{GibbonsFreedman} D. Z. Freedman and G. W. Gibbons, {\sl Remarks on
supersymmetry and K\"ahler geometry}, in {\it Superspace and
Supergravity}, eds S.W. Hawking and M. Rocek (Cambridge University
Press, (1981)) 449-454; see also
D.Z. Freedman and  G.W. Gibbons, {\sl New higher dimensional Ricci flat 
K\"ahler metrics}, ITP-SB-8 0-48 (1980).

\bibitem{Townsend} P.K. Townsend, {\sl $p$-brane democracy},
PASCOS/Hopkins 1995, hep-th/9507048.

\bibitem{cglpg2} M.~Cveti\v c, G.W. Gibbons, H.~L\"u and  C.~N.~Pope,
{\sl Supersymmetric M3-branes and $G_2$ manifolds},
hep-th/0106026.

\bm{cglpsten} M. Cveti\v{c}, G.W. Gibbons, H. L\"u and C.N. Pope, {\sl
Ricci-flat metrics, harmonic forms and brane resolutions}, hep-th/0012011.

\bm{candel} P. Candelas and X.C. de la Ossa, {\sl Comments on
conifolds},  Nucl. Phys. {\bf B342}, 246 (1990). 

\bm{Onemli-Tekin} V.K. Onemli and B. Tekin, {\sl Kaluza-Klein vortices}, 
JHEP {\bf 0101}, 034 (2001), hep-th/0011287.

\bm{Stelle} G.W. Gibbons, H. L\"u, C.N. Pope and K.S. Stelle,
{\sl Supersymmetric domain walls from metrics of special holonomy},
hep-th/0108191. 

\bibitem{Lorenz}  D. Lorenz-Petzold, {\sl Positive-definite self-dual 
solutions of Einstein's field equations},  J. Math. Phys. {\bf 24},
2632 (1983); {\sl Gravitational instanton solutions}, Prog.
Theor. Phys. Letters {\bf 81}, 17 (1989); 
D. Lorenz,  {\sl Gravitational instanton solutions for Bianchi Types I-IX},
Acta Physica Polonica {\bf B14}, 791 (19xx).

\bm{begipapo} V.A. Belinsky, G.W. Gibbons, D.N. Page and C.N. Pope,
{\sl Asymptotically Euclidean Bianchi IX Metrics In Quantum Gravity,}
Phys. Lett. {\bf B76}, 433 (1978).

\bm{lalupo} I.~V.~Lavrinenko, H.~L\"u and C.~N.~Pope,
{\sl Fibre bundles and generalised dimensional reductions},
Class.\ Quant.\ Grav.\  {\bf 15}, 2239 (1998), hep-th/9710243. 

\bibitem{gibryc} G.~W.~Gibbons and P.~Rychenkova,
{\sl Single-sided domain walls in M-theory},
J.\ Geom.\ Phys.\  {\bf 32}, 311 (2000), hep-th/9811045.

\bibitem{Page1} D.N. Page, {\sl A periodic but nonstationary gravitational
instanton},  Phys. Lett. {\bf B100}, 313  (1981).

\bm{Hitchin1} N.J. Hitchin, {\sl Twistor construction of Einstein metrics}, 
in {\it Global Riemannian Geometry} N.J. Hitchin and T. Wilmore eds. 
(Ellis Horwood, Chichester (1984)).

\bm{Kobayashi1} R. Kobayashi, {\sl Moduli of Einstein metrics on a K3
surface and degenerations of type I}, Advanced Studies in
Mathematics {\bf 18-II}, 257 (1990).

\bibitem{schwarz} A.S. Schwarz, {\sl Field theories with no local
conservation of the electric charge}, Nucl. Phys. {\bf B208}, 141 (1982).

\bm{hawk} S.W. Hawking, {\sl Gravitational instantons}, Phys. Lett. {\bf
A60}, 81 (1977). 

\bm{gibhaw} G.W. Gibbons and S.W. Hawking, {\it Gravitational
multi-instantons}, Phys.\ Lett.\ B {\bf 78}, 430 (1978). 

\bibitem{Gibbons-Pope} G.W. Gibbons and C.N. Pope, {\sl The positive 
action conjecture and asymptotically Euclidean metrics in quantum 
gravity}, Comm. Math. Phys. {\bf 66}, 267 (1979).

\bibitem{Page2} D.N. Page, {A physical picture of the
K3 gravitational instanton}, Phys. Lett. {\bf 80 B}, 55 (1978).

\bm{Kobayashi2} R. Kobayashi, {\sl Ricci-flat K\"ahler metrics
on affine algebraic manifolds and degenerations of K\"ahler-Einstein
K3 surfaces},  Advanced Studies in Pure Mathematics {\bf 18-II}, 137 
(1990).

\bm{Nergiz-Saclioglu} S. Nergiz and C. Saclioglu, {\sl A quasiperiodic
Gibbons-Hawking metric and space-time foam}, Phys. Rev. {\bf D53}, 2240 
(1996),  hep-th/9505141.

\bm{Anderson-Kronheimer-LeBrun} M. Anderson, P. Kronheimer and C.
LeBrun, {\sl Complete Ricci-flat K\"ahler manifolds of infinite
topological type}, {\it Comm. Math. Phys.} {\bf 125}, 637  (1989).

\bm{Ooguri-Vafa} H. Ooguri and C. Vafa, {\sl Summing up D instantons}, 
Phys. Rev. Lett. {\bf 77}, 3296 (1996), hep-th/968079.

\bm{Greene} B.R. Greene, A.D. Shapere, C. Vafa and S.T. Yau, {\sl
Stringy cosmic strings and non-compact Calabi-Yau manifolds}, 
Nucl. Phys. {\bf B337}, 1 (1990). 

\bm{Gross-Wilson} M. Gross and P.M.H. Wilson, {\sl Large complex structure
limit of K3 surfaces},\newline 
math.DG/0008018.

\bm{gigrpe} G.W. Gibbons, M.B. Green and M.J. Perry, {\sl Instantons 
and seven-branes in type IIB Superstring Theory}, Phys. Lett. 
{\bf B370}, 37  (1996), hep-th/9511080

\bibitem{hawpop} S.W. Hawking and C.N. Pope, {\sl Symmetry breaking by
instantons in supergravity}, Nucl. Phys. {\bf B146}, 381 (1978). 

\bm{Ward} R. Ward, {\sl Einstein-Weyl spaces and $SU(\infty)$ Toda fields},
Class. Quantum Grav. {\bf 7}, L95 (1990).

\bm{Calderbank-Tod} D.M.J. Calderbank and P. Tod, {\sl
Einstein metrics, hypercomplex structures and the Toda field equation},
Diff. Geom. Appl. {\bf 14}, 199 (2001), math.DG/9911121.

\bm{Nutku-Sheftel} Y. Nutku and M.B. Sheftel, {\sl A family of 
heavenly metrics}, gr-qc/0105088.

\bm{Bakas-Sfetsos} I. Bakas and K. Sfetsos, {\sl Toda fields of $SO(3)$ 
hyper-K\"ahler metrics}, Nucl. Phys. Proc. Suppl. {\bf 49}, 10 (1996), 
hep-th/9601087; {\sl Toda fields of $SO(3)$ 
hyper-K\"ahler metrics and free field realizations}, Int. J. Mod. Phys. 
{\bf A12}, 2585 (1997), hep-th/9604003.

\bibitem{GibbonsManton} G.W. Gibbons and N.S. Manton, {\sl The moduli space
metric for well separated BPS monopoles}, Phys. Lett. {\bf B356}, 32 
(1995), hep-th/9506052.


\end{thebibliography}
\end{document}